\documentclass[lettersize,journal, onecolumn]{IEEEtran}
\usepackage{amsmath,amsfonts}
\usepackage{algorithmic}
\usepackage{algorithm}
\usepackage{array}
\usepackage[caption=false,font=normalsize,labelfont=bf,textfont=sf]{subfig}
\usepackage[labelfont=bf]{caption}
\usepackage{textcomp}
\usepackage{stfloats}
\usepackage{float}
\usepackage{url}
\usepackage{verbatim}
\usepackage{graphicx}
\usepackage{cite}
\usepackage{siunitx}
\usepackage{balance}
\usepackage{circuitikz}
\usepackage{capt-of}
\usepackage[export]{adjustbox}
\usepackage{subfig}
\usepackage{hyperref}
\usepackage{color,soul}
\usepackage{titlesec}
\usepackage{orcidlink}
\usepackage{xr}
\usepackage{makecell}
\usepackage{mhchem}

\titleformat*{\section}{\large\bfseries}
\titleformat*{\subsection}{\bfseries}

\newcommand{\ignore}[1]{}

\newdimen\OneColumnWidth                 
\newdimen\OneColumnWidthWide                 
\newdimen\TwoColumnWidth                    
\begin{document}
\bstctlcite{IEEEexample:BSTcontrol}
	\title{Spinwave Bandpass Filters for 6G Communication}
    % Spinwave Bandpass Filter for 6G
    % Scalable Spinwave Bandpass Filters
	
	\author{Connor Devitt$^{1\ast}$ \orcidlink{0000-0002-3394-6842}, Sudhanshu Tiwari$^{1}$ \orcidlink{0000-0002-5499-0877}, Bill Zivasatienraj$^{2}$ \orcidlink{https://orcid.org/0000-0002-0522-4869}, Sunil A. Bhave$^{1\ast}$ \orcidlink{0000-0001-7193-2241}
		% <-this % stops a space
            \\ $^{1}$OxideMEMS Lab, Elmore Family School of Electrical and Computer Engineering, Purdue University, West Lafayette, IN 47907 USA. 
            % \\ $^{2}$Amazon Web Services, Pasadena, CA 91106.
            \\ $^{2}$FAST Labs\textsuperscript{TM}, BAE Systems, Inc., Nashua, NH 03060 USA. 
            \\ $^\ast$e-mail: devitt@purdue.edu, bhave@purdue.edu
        }

	% The paper headers
	% \markboth{}%
	{}
	\maketitle

\section*{Abstract}
Spinwave filters using single-crystal yttrium iron garnet are an attractive technology for integration in frequency adjustable or tunable communication systems. However, existing SW devices do not have sufficient bandwidth for future 5G and 6G communication systems, are too large, or have strong spurious passbands creating unintentional cross-channel interference. Leveraging modern micromachining fabrication methods capable of wafer-scale production, we report a SW ladder filter architecture requiring only a single external magnetic bias. The filters demonstrate loss as low as $\SI{2.54}{\decibel}$, bandwidths up to $\SI{663}{\mega\hertz}$, center frequency tuning over multiple octaves from $7.08-21.6 \text{ GHz}$, and high linearity with an input referred 3\textsuperscript{rd}-order intercept point over $11 \text{ dBm}$ in the passband. The filter's operation is also experimentally demonstrated in a frequency tunable radio system.
 
\section*{Introduction}
    
    Modern wireless communication systems are omnipresent supporting everything from mobile smartphones, internet of things (IoT) devices, and WiFi to satellite communication and radar. In pursuit of higher data rates, power efficiency, and spectral efficiency, next generation communication systems including the FR3 band for 5G at $\SI{7.125}{\giga\hertz}$ -- $\SI{24.25}{\giga\hertz}$ and 6G \cite{wang_road_2023, andrews_6g_2024} are pushing for higher channel frequencies and larger bandwidths. Radio frequency (RF) front end modules are critical subsystems present in every transceiver system consisting of (at minimum) an antenna, bandpass filter, low-noise amplifier or power amplifier, mixer, and local oscillator (LO). Improving the performance of the bandpass filter in the RF front end is crucial to realize RF front modules at the high channel frequencies required by next generation communication systems. The filter performance must be exceptional showing low insertion loss to preserve high signal-to-noise ratio (SNR) at the receiver and maintain high power efficiency at the transmitter. The bandwidth should be sufficiently large ($100$s MHz) to support high data rates while strongly rejecting interference from neighboring channels. Additionally, the filter must demonstrate high power handling and linearity within in compact footprint and be manufacturable using a high yield fabrication process. \\

    Below $\SI{6}{\giga\hertz}$, microacoustic filters based on surface acoustic waves (SAW) and bulk acoustic waves (BAW) dominate \cite{hagelauer_microwave_2023, ruby_snapshot_2015} the filter market primarily due to their low loss, micro-scale footprint, and mature fabrication process that is capable of producing billions of units per year with high yield. Pushing acoustic wave filters to higher frequencies with large bandwidths is an active research area with recent advances demonstrating filters in $\SI{7}{\giga\hertz}$ to $\SI{24}{\giga\hertz}$. Advanced materials such as thin-film lithium niobate (\ce{LN}) \cite{anusorn_frequency_2025, yang_sv-saw_2025} and scandium doped aluminum nitride (\ce{ScAlN}) \cite{izhar_periodically_2025, giribaldi_compact_2024}, in combination with new fabrication processes and design techniques have enabled low loss and wide band acoustic filters at high frequencies.\\

    Commercial RF front end systems, in a mobile phone for example, require more than one hundred acoustic filters to cover every Tx and Rx frequency band \cite{emilio_resonants_2021}. The number of frequency bands (and the number of filters) required to cover the FR3 band and future 6G bands is expected to grow \cite{ruby_snapshot_2015} incurring significant increases to system costs and size. Therefore, there is a strong desire to develop reconfigurable filtering solutions \cite{hagelauer_microwave_2023} that could allow a single filter to cover multiple frequency bands. Reconfigurable acoustic filters can be constructed using ferroelectric materials such as barium strontium titanate (BST) \cite{zhu_dc_2007}, phase change materials \cite{hummel_reconfigurable_2015, fouladi_azarnaminy_switched_2022}, and varactors \cite{hashimoto_moving_2015}. A natural, frequency tunable alternative, however, is to use magnetic materials in the component resonators instead of piezoelectric or ferroelectric materials.\\

    In ferromagnetic or ferrimagnetic films such as yttrium iron garnet (YIG), spinwaves are propagating disturbances in the film's precessing magnetization \cite{stancil_theory_1993}. The dispersion of SW (Supplementary Information \ref{sup-spinwave_dispersion}) is highly nonlinear and can be tuned by adjusting the film's external magnetic bias \cite{ishak_magnetostatic_1988}. A filter's footprint is primarily determined by the wavelength. For microacoustic devices, their slow group velocities and short wavelengths compared to electromagnetic waves provide significant device miniaturization. The wavelengths of spinwaves, in comparison, are positioned between acoustic waves and electromagnetic waves (Fig. S\ref{sup-SI_Fig_MSFVW_dispersion}). Therefore, in addition to their inherent tunability, one major advantage of SW is their ability to miniaturize devices independently of the operating frequency \cite{levchenko_review_2024}. However, unlike microacoustic resonators whose key figures of merit such as quality factor ($Q$-factor) and effective coupling ($k_{eff}^2$) reduce at high frequency, the corresponding figures of merit for SW resonators improve at higher frequencies \cite{tiwari_high-performance_2025}. These attractive features inherent to SW coupled with the increasing need for tunable RF devices have created a resurgence in SW research. Recent advances in microfabrication and packaging techniques for thin film magnetic materials \cite{tiwari_high-performance_2025, du_frequency_2024, du_wideband_2025, devitt_distributed_2024, wang_temperature_2024, tikhonov_temperature_2013} have reduced the device footprint and have enabled fabrication at the chip and wafer scale, which is a requirement for mass production -- a significant step forward compared to large YIG sphere devices \cite{micro_lambda_wireless_inc_mlfd_nodate} or bulk YIG flip-chip devices \cite{marcelli_tunable_1991, adam_msw_1985, yang_low-loss_2013, wu_nonreciprocal_2012, tsai_wideband_2009, zhu_tunable_2009}.\\

    The channel bandwidths for 5G FR3 and future 6G bands are expected to be $100-400 \text{ MHz}$ \cite{holma_extreme_2021, giribaldi_compact_2024} and potentially over $\SI{1}{\giga\hertz}$ \cite{anusorn_frequency_2025, shakya_comprehensive_2024} so the bandpass filters must have correspondingly large bandwidths to support these channels. However, previously reported SW and magnetostatic wave (MSW) filters do not have sufficiently high bandwidths, are too large, or have strong spurious modes leading to unintentional interference between neighboring channels (see related work in \cite{wu_spatially_2025}). In this article, we present a compact SW ladder filter comprised of high figure-of-merit SW resonators based on \cite{tiwari_high-performance_2025}. The ladder filter is a robust architecture used by billions of acoustic filters each year, but requires two resonators with a significant resonance frequency shift. For SW, this shift would typically require two separate uniform external magnetic fields resulting in a larger packaged filter volume, increased filter routing loss, and added complexity and cost. This work develops a method to realize a resonance frequency shift using only a single external magnetic bias based on a deep Ar ion etching microfabrication technique. Both 3\textsuperscript{rd}-order and 5\textsuperscript{th}-order ladder filters are demonstrated with lithographically adjustable filter bandwidths as high as $\SI{663}{\mega\hertz}$, low loss, strong spurious suppression, high linearity, and multi-octave center frequency tuning over the $5.7-21.6 \text{ GHz}$ range. The filters use an optimized fabrication process with high device yield over a $15\times15 \text{ mm$^2$}$ chip which can easily be adapted for wafer-scale fabrication. By requiring only one external magnetic field, the volume of the SW filter and its packaged magnetic bias is substantially reduced while simultaneously achieving higher performance. Furthermore, the packaging can leverage recently reported micromagnetic assemblies \cite{wang_temperature_2024, du_frequency_2024, tikhonov_temperature_2013}. As an application demonstration of this technology, we show a frequency agile quadrature amplitude modulated (QAM) radio receiver that shows good immunity to nearby interference.

\section*{Spinwave Ladder Filter Design}
    \begin{figure}[!b]
        \centering
        \includegraphics[width=\textwidth]{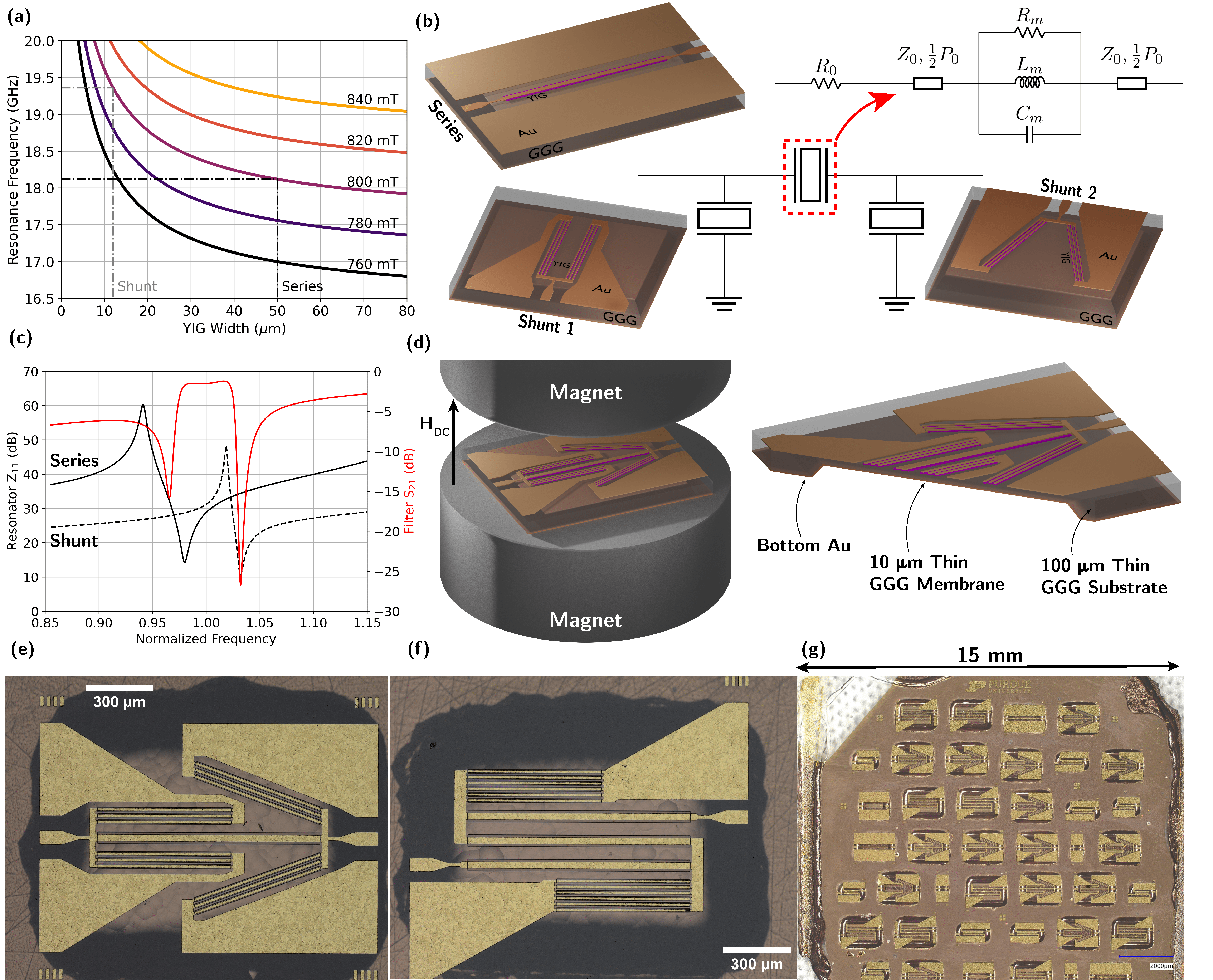}
        \caption{\textbf{Design of SW ladder filters with a single magnetic bias.} \textbf{(a)} Dispersion curve for magnetostatic forward volume waves (MSFVW) including demagnetization effects over the YIG resonator width for different magnetic bias fields. With high contrast in the YIG dimensions between the series and shunt resonators, a large resonance frequency separation can be realized using a single external magnetic bias. \textbf{(b)} 3-D schematic of series and shunt SW resonators in a ladder filter configuration. Each resonator in the filter schematic is described by a 2-port distributed SW model based on \cite{devitt_distributed_2024}. The series resonator consists of a single, wide YIG mesa ($\SI{1000}{\micro\meter}\times \SI{50}{\micro\meter} \times \SI{3}{\micro\meter}$) and a high impedance Au transmission line. The two shunt resonators are constructed from an array of parallel YIG fins ($\SI{600}{\micro\meter}\times \SI{12}{\micro\meter} \times \SI{3}{\micro\meter}$) with a low input impedance. \textbf{(c)} Modeled SW resonator impedance response and resulting 3\textsuperscript{rd}-order filter response of the cascaded series and shunt resonators. The resonator models are fit from finite element simulations (Ansys HFSS) at $\SI{792.3}{\milli\tesla}$ bias field. \textbf{(d)} Rendering of a 3\textsuperscript{rd}-order SW ladder filter layout consisting of 1 series YIG resonator and 12 YIG fins collectively acting as 2 shunt resonators. The filter is situated between two permanent magnets providing the out-of-plane magnetic bias ($H_{DC}$). A cross section of the filter layout highlights the $\SI{10}{\micro\meter}$ thin GGG membrane and Au ground plane beneath each device. \textbf{(e)} fabricated 3\textsuperscript{rd}-order and \textbf{(f)} 5\textsuperscript{th}-order SW ladder filters. The dark region surrounding each filter is the sidewall of the etched GGG cavity beneath each device. \textbf{(g)} Chip micrograph showing a multitude of functional SW ladder filters, resonators, and de-embedding structures.}
        \label{Fig_Design}
        % \vspace*{-0.1in}
    \end{figure}

    The design and operation of SW ladder filters can be understood through an analogy to microelectromechanical system (MEMS) acoustic filters \cite{giribaldi_620_2023}. They are constructed from alternating series and shunt piezoelectric resonators whose frequency responses are described using a modified Butterworth-Van Dyke (mBVD) model \cite{larson_modified_2000} (see Supplementary Information \ref{sup-Acoustic_analogy}). The piezoelectric resonators use a capacitive transducer to excite a particular acoustic mode. They exhibit both a high admittance series resonance at $f_s$ corresponding to the mechanical mode and a low admittance parallel anti-resonance at $f_p$. The filter passband is formed by introducing a frequency shift such that $f_s$ of the series resonator aligns with $f_p$ of the shunt resonator giving a low impedance path from port 1 to port 2 and a high impedance path to ground (Fig. S\ref{sup-SI_Fig_Acoustic_SW_compare}) \cite{hashimoto_rf_2009}. Far away from the mechanical resonance, the MEMS acoustic filter behaves as a capacitive divider network. Therefore, the out-of-band rejection is determined by the impedance contrast of the capacitive transducers as well as the number of resonator stages in the filter.

    The transduction of spinwaves relies on a transverse RF magnetic field instead of an electric field in the case of piezoelectric resonators. Therefore, the electrical response of the SW transducer is modeled using a series $R_0$ and an inductive transmission line with impedance $Z_0$ and physical length $P_0$. The spinwave resonance is modeled as a parallel resonator network ($R_m$, $C_m$, $L_m$) in series with the transducer as shown in Fig. $\ref{Fig_Design}$b \cite{devitt_distributed_2024}. Far away from resonance, the spinwave ladder filter behaves as an inductive divider network. The out-of-band rejection is a function of the number of resonators stages and the impedance contrast between the series and shunt transmission lines. This distributed SW resonator model exhibits a high impedance resonance at $f_p$ corresponding to the selected SW mode and a low impedance anti-resonance at $f_s$. To realize a bandpass filter, the shunt resonator's $f_p$ must be shifted to roughly align with $f_s$ of the series resonator. The resonators in this work rely on forward volume spinwave modes which exist when using an out-of-plane magnetic bias. The dispersion of these modes is a function of the effective magnetic field inside the film ($H_{DC}^{eff}$), the film thickness $t_{YIG}$, and the spinwave vector $k_{mn}$ which depends on the film geometry \cite{ishak_magnetostatic_1988, stancil_theory_1993}. A significant shift between the series and shunt resonances can be accomplished by biasing each with a distinct external magnetic field ($H_{DC}$). However, realizing two strong, uniform magnetic biases over each SW resonator in a small filter area is exceptionally difficult and would increase the packaged filter volume. Furthermore, the physical separation between the series and shunt resonators would need to increase resulting in higher routing losses and reduced center frequency tuning range. Alternatively, engineering a contrast in the spinwave wave vectors to obtain the frequency shift required for filters with even a moderate bandwidth is equally as challenging due to the extremely nonlinear spinwave dispersion (Fig. S$\ref{sup-SI_Fig_MSFVW_dispersion}$). The effective bias is given by $H_{DC}^{eff}=H_{DC}-N_zM_s$ where $N_zM_s$ is the film's demagnetizing field in the out-of-plane direction and $M_s=\SI{178}{\milli\tesla}$ is the saturation magnetization for YIG. The demagnetization factor $N_z$ is strongly geometry dependent and can be calculated using \cite{aharoni_demagnetizing_1998} for rectangular prisms. By engineering a contrast in the series and shunt resonator dimensions, this work combines the frequency shift due to the variation in $k_{mn}$ with the shift due to the demagnetization field to realize filters with over $\SI{600}{\mega\hertz}$ in bandwidth biased by a single external magnetic field. Fig. $\ref{Fig_Design}$a illustrates the SW resonance frequency over the resonator width (with a nominal length and thickness of $\SI{600}{\micro\meter}$ and $\SI{3}{\micro\meter}$ respectively) for a range of external bias fields, $H_{DC}$. With a shunt resonator width of $\SI{12}{\micro\meter}$ and a series resonator width of $\SI{50}{\micro\meter}$, the calculated resonance frequency separation is $\SI{1.24}{\giga\hertz}$ due to the combined contrast in $k_{mn}$ and $N_z$.\\

    Designing ladder filters with appreciable bandwidth and low loss also requires the component resonators to have high quality factors ($Q$-factors) at both $f_p$ and $f_s$ and sufficiently high coupling factor, $k_{eff}^2$, (defined in Supplementary Information \ref{sup-resonator_design_modeling}). Widely available single crystal YIG films grown on lattice matched gadolinium gallium garnet (GGG) substrates using liquid phase epitaxy (LPE) gives high $Q$-factors at $f_p$ due to the material's low inhomogeneous broadening and low Gilbert damping factor on the order of $10^{-4}$ \cite{gilbert_phenomenological_2004, chumak_advances_2022, beaujour_ferromagnetic_2009}. Recently reported experimental $Q$-factors in this material platform include $Q=1313$ at $f_p=\SI{11.6}{\giga\hertz}$ \cite{du_frequency_2024}, $Q=2206$ at $f_p=\SI{6.79}{\giga\hertz}$ \cite{devitt_edge-coupled_2024}, and $Q=200-350$ at $f_p=4-11 \text{ GHz}$ \cite{costa_compact_2021}. The $Q$-factor at $f_s$ is primarily a function of the transmission line design and can be maximized using low resistivity Au transducers \cite{devitt_distributed_2024}. Until recently, effective resonator coupling has been limited to $k_{eff}^2<3\%$ \cite{du_meander_2024, costa_compact_2021, feng_micromachined_2023} using conventional fabrication techniques that allow only topside electrodes for YIG-on-GGG resonators. The effective coupling for SW resonators strongly depends on the confinement of the transverse RF magnetic field within the YIG film \cite{tiwari_high-performance_2025}. To push the experimentally demonstrated $k_{eff}^2$ up to $18\%$, \cite{tiwari_high-performance_2025} introduces a backside anisotropic GGG wet etch process to place a ground plane in close proximity ($<\SI{20}{\micro\meter}$) below the YIG film with thru-GGG vias to wrap the transducer around the YIG resonator. \\
    
    Building off of the high coupling resonators in  \cite{tiwari_high-performance_2025}, Fig. $\ref{Fig_Design}$b shows a rendering of the series and shunt component resonators for the SW ladder filter. The series resonator consists of a $\SI{1000}{\micro\meter}\times \SI{50}{\micro\meter} \times \SI{3}{\micro\meter}$ ion-milled YIG mesa with a bottom ground plane $\SI{10}{\micro\meter}$ beneath the YIG film. The transducer uses $\SI{3}{\micro\meter}$ electroplated gold to minimize resistance ($R_0$). Based on the study in \cite{devitt_distributed_2024}, the transducer width is $60\%$ of the YIG width to strongly suppress spurious SW modes. The shunt resonator consists of a parallel array of $6$ narrow YIG fins each measuring $\SI{600}{\micro\meter}\times \SI{12}{\micro\meter} \times \SI{3}{\micro\meter}$. The relatively narrow shunt resonator reduces the demagnetization field and increases $k_{mn}$, resulting in a higher resonance frequency at the same externally applied bias. Using 6 YIG fins in parallel provides a balance between a low off-resonance impedance for strong out-of-band rejection and a high shunt resonator $Q$-factor for low insertion loss.  Consistent with the series resonator, the $\SI{3}{\micro\meter}$ thick Au shunt transducer is $60\%$ of the YIG width and a ground plane is $\SI{10}{\micro\meter}$ away. Fig. $\ref{Fig_Design}$c illustrates the SW ladder filter operation by plotting the input impedance of the series and shunt resonators against the modeled filter response for a 3\textsuperscript{rd}-order 1-series, 2-shunt SW ladder filter. Further details on resonator and filter design, modeling, and measured performance are available in Supplementary Information \ref{sup-resonator_design_modeling}. Specifically, it introduces a method to tune the distributed resonator model in Fig. $\ref{Fig_Design}$b to predict the resonator and ladder filter frequency response over magnetic field tuning.  Furthermore, it discusses the resonator and filter performance trends over both the YIG width and the distance between the YIG and bottom ground plane. Fig $\ref{Fig_Design}$d illustrates a 3\textsuperscript{rd}-order SW ladder filter layout by combining the individual series and shunt resonators from Fig $\ref{Fig_Design}$b. The filter is situated between two magnets that generate a single external out-of-plane magnetic bias. A cross section of the filter layout highlights the selectively thinned GGG substrate enabling the placement of a close-proximity ground plane beneath the filter (required for high resonator $k_{eff}^2$). To experimentally validate these SW ladder design concepts, they are fabricated using a $15\times 15 \text{ mm}$ YIG-on-GGG chip. Fig. $\ref{Fig_Design}$e and Fig. $\ref{Fig_Design}$f show micrographs of a 3\textsuperscript{rd}-order and 5\textsuperscript{th}-order fabricated SW filter respectively. Fig. $\ref{Fig_Design}$g highlights the complete YIG-on-GGG chip filled with a multitude of functional filters, resonators, and de-embedding structures.\\

    Since the frequency separation between the component resonators is geometry dependent, the resulting SW ladder filter bandwidth can be lithographically tuned by adjusting the width of the shunt YIG resonators (changing the contrast in the demagnetization field). Furthermore, since the frequency separation is insensitive to the external magnetic bias, the filters tune with nearly constant absolute bandwidth. Based on the simulated and measured resonator performance (Supplementary Information \ref{sup-resonator_design_modeling}), filter bandwidths in excess of $\SI{1}{\giga\hertz}$ are possible but there is an inherent trade-off between the filter's center frequency tuning range and bandwidth. At low frequencies, the resonator's $k_{eff}^2$ does not increase fast enough to support the filter's rapidly increasing fractional bandwidth, resulting in a large passband ripple. At high frequencies, the SW transducers become electrically long and the filter stops behaving as an inductive divider network, resulting in poor filter rejection. This upper center frequency tuning limit can be extended by reducing physical length of each resonator at the expense of lower resonator $k_{eff}^2$, and by extension, filter bandwidth. A lower center frequency tuning limit can be attained by reducing the designed filter bandwidth (using wider shunt YIG resonators for example) or the resonator coupling can be increased by reducing the GGG membrane thickness ($t_{GGG}$) beneath each filter at the cost of resonator $Q$-factor and filter insertion loss. Moderate filter bandwidths in the range of $100-400 \text{ MHz}$ are easier to design and fabricate with this technology and show superior loss and tuning range due to the reduced resonator effective coupling requirements.

\section*{Experimental Results}
    \begin{figure}[!t]
        \centering
        \includegraphics[width=\textwidth]{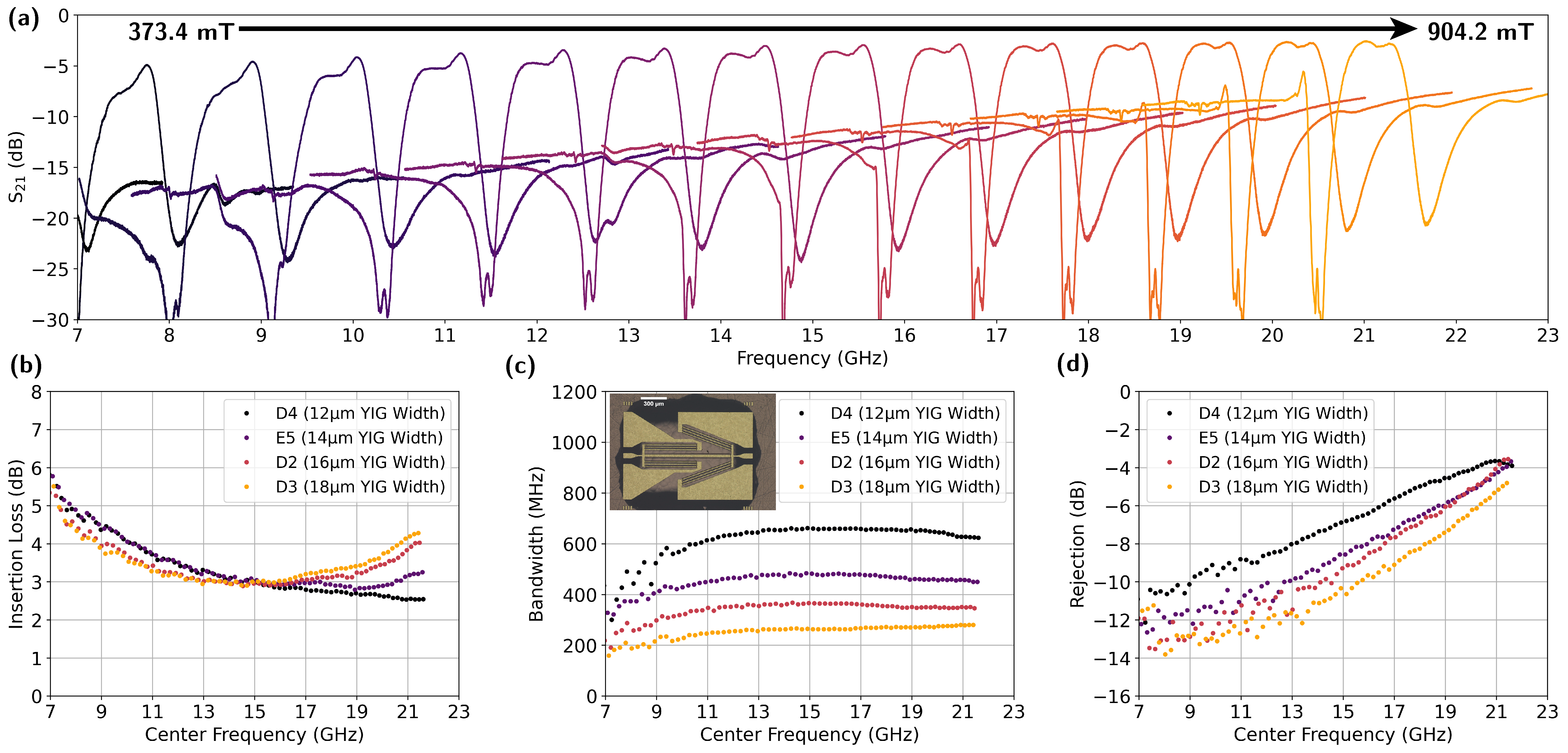}
        \caption{\textbf{Measured SW filter frequency response.} The S-parameters of the 3\textsuperscript{rd}-order filter (Device D4) shown in Fig. $\ref{Fig_Design}$c are measured over magnetic bias using the experimental setup detailed in Supplementary Information \ref{sup-Measurement_Setup}. The probe pads are de-embedded from the filter measurements using device D5 and the port impedance is renormalized to $Z_0=\SI{15}{\ohm}$. \textbf{(a)} SW ladder filter transmission response with an out-of-plane bias ranging from $\SI{373.4}{\milli\tesla}$ to $\SI{904.2}{\milli\tesla}$. The response is cropped around the passband for clarity. \textbf{(a)-(c)} The filter's measured insertion loss (IL), $\SI{3}{\decibel}$ Bandwidth (BW), and rejection level respectively over magnetic bias and shunt resonator YIG width. The reported rejection is the worst-case $S_{21}$ 5 bandwidths away from the center frequency relative to the filter's insertion loss. The raw S-parameters are available alongside this paper in a Zenodo database.}
        \label{Fig_filterResponse}
        % \vspace*{-0.1in}
    \end{figure}

    \begin{table} [!b]
        \caption{Performance Comparison with Other Tunable Bandpass Filters}
        \centering
        \renewcommand{\arraystretch}{1.5}
        \begin{tabular}{|p{2.5cm}|p{1.6cm}|p{1.3cm}|p{1.8cm}|p{1.5cm}|p{2.9cm}|p{1.6cm}|p{1.6cm}|}\hline
            Reference  & Frequency \hspace{0.1cm} Tuning (GHz)  & Insertion Loss (dB) & Bandwidth & Rejection (dB) & IIP3 (dBm) & P1dB (dBm)\\\hline
            This work 3\textsuperscript{rd}-order & 7.08--21.6 & 2.54--5.78 & 281--663 & 3.64--13.81 & $>11$ at \SI{16}{\giga\hertz} with $\SI{1}{\mega\hertz}$ tone spacing & $>$5 \hspace{1cm} at \SI{16.9}{\giga\hertz}\\\hline
            This work 5\textsuperscript{th}-order & 7.44--20.36 & 6.81--11.66 & 255--539 & 14.61--24.82 & -- & --\\\hline
            YIG MSFVW \cite{devitt_edge-coupled_2024} & 4.5--10.1 & 11 & 11--17 MHz & $35$ & $-4.85$ at \SI{7.6}{\giga\hertz} with $\SI{15}{\mega\hertz}$ tone spacing & -- \\\hline
            YIG MSSW \cite{du_frequency_2024}  & 3.4--11.1 & 3.2--5.1 & 18--25 MHz & $25$ & $-8.36$ at \SI{5.80}{\giga\hertz} with  $\SI{1}{\mega\hertz}$ tone spacing & -19.86 \hspace{1cm} at $\SI{6.2}{\giga\hertz}$ \\\hline
            YIG MSSW \cite{du_wideband_2025}  & 4.0--17.5 & 2.9--4.9 & 107--198 MHz & $30$ & -- & 12.4 \hspace{1cm} at \SI{9.1}{\giga\hertz} \\\hline
            YIG MSSW \cite{wu_spatially_2025}  & 6.3--16.8 & 2.4--3.2 & 181-211 MHz & 25--39 & -- & -- \\\hline
            YIG \cite{tsai_wideband_2009} & 5--19 & 1.7--2 & 62--107 MHz & $20$  & -- & -- \\\hline
            RF MEMS \cite{entesari_differential_2005} & 6.5--10 & 4.1--5.6 & 306--539 MHz & $50$  & $>45$ at \SI{9.80}{\giga\hertz} with  $\SI{0.5}{\mega\hertz}$ tone spacing & 24$^\ast$ \\\hline
            65 nm CMOS \cite{liu_1017_2025} & 10--17 & 3--4 & 3.9\% - 8.8\%  & $35$  & -- & -13.6 -- -6 \\\hline
            \multicolumn{7}{p{14.6cm}}{$^\ast$ Self actuation power} \\
        \end{tabular}
        %		\vspace{-0.15in}
        \label{Performance_Comparison}
    \end{table}

    \begin{figure}[!b]
        \centering
        \includegraphics[width=\textwidth]{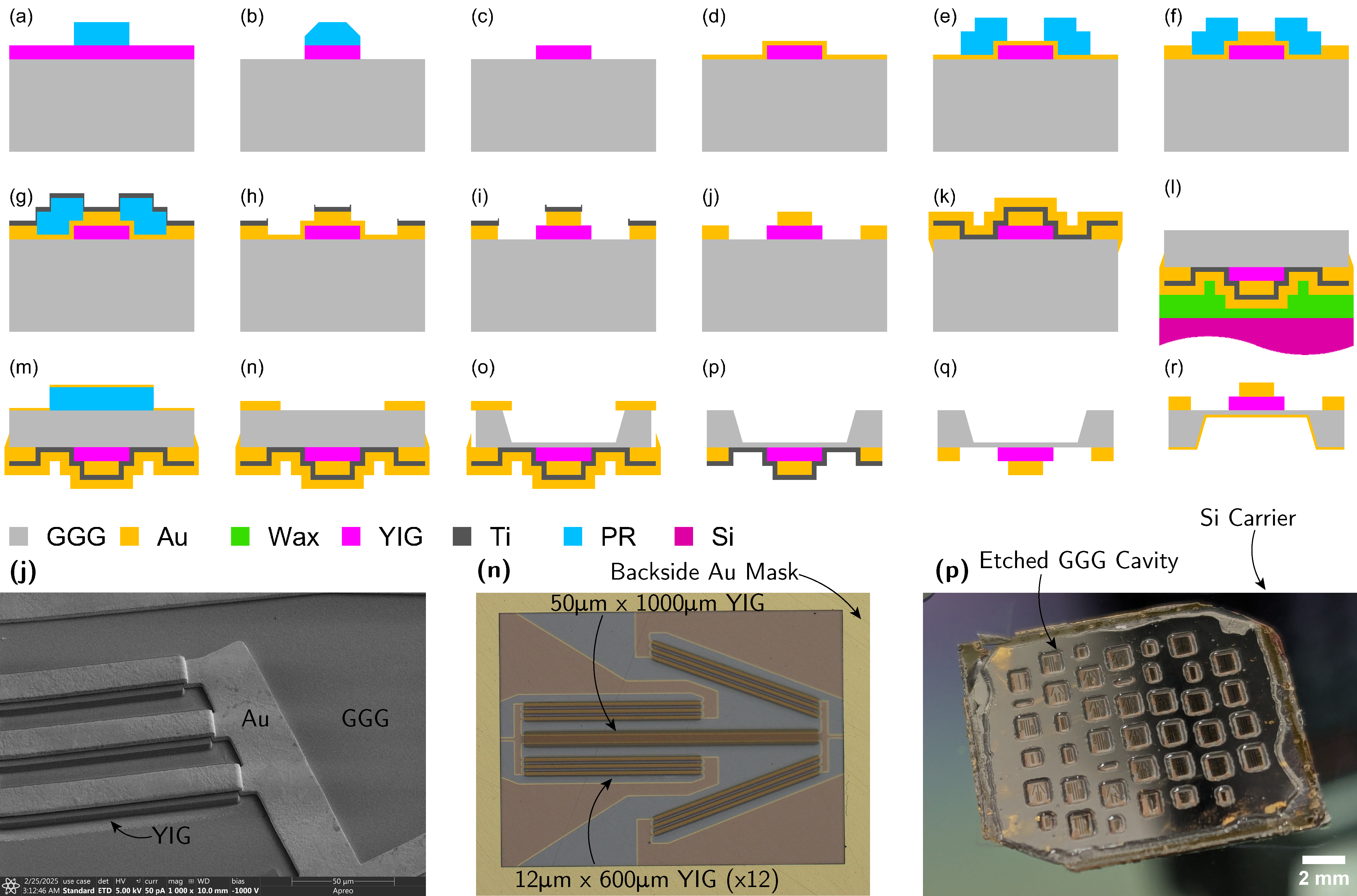}
        \caption{\textbf{SW ladder filter fabrication process.} Steps \textbf{(a)-(j)} based on \cite{devitt_distributed_2024} show the YIG ion milling and patterned electroplating processes which create the final YIG resonators and thick gold transducers. The thickness of the Ti hard mask (for the the Au etch in step $\textbf{(i)}$) is $\SI{150}{\nano\meter}$ while the Ti adhesion layer in $\textbf{(d)}$ is $\SI{15}{\nano\meter}$ thick. This ensures the narrow shunt transducers are not completely undercut during the buffered oxide etchant (BOE) Ti wet etch in step \textbf{(j)}. A SEM of one the ladder filter's shunt resonators is shown after step \textbf{(j)}. Steps \textbf{(k)-(r)} represent a simplified version of the hairclip YIG resonator fabrication from \cite{tiwari_high-performance_2025} without thru vias and optimized for higher yield and metal quality. \textbf{(k)} A $300/100/785 \text{ nm}$ Ti/Au/Electroplated Au layer is used to protect the YIG resonators and Au transducers against a hot phosphoric acid etch in step \textbf{(o)}. The chip is bonded to a Si carrier chip using a thin spun-on wax and mechanically polished down to \SI{101}{\micro\meter} in \textbf{(l)}. The wax process is optimized to ensure the YIG-on-GGG chip is flat relative to the Si carrier giving a thickness uniformity within $\pm 1 \si{\micro\meter}$ after thinning. Steps \textbf{(m)-(n)} define a backside electroplated Au hard mask through a liftoff process. A micrograph of a SW ladder filter, taken after step \textbf{(n)}, shows the electroplated Au hard mask viewed from the backside of the chip. The image also highlights the undercutting of the Ti adhesion layer at the edges of the frontside transducers caused by the BOE etch in step \textbf{(j)}. As detailed in \cite{tiwari_high-performance_2025}, $160^\circ C$ phosphoric acid is used to etch the GGG substrate leaving a $\SI{10}{\micro\meter}$ membrane beneath each device. In \textbf{(p-r)}, the Au and Ti hard masks are removed via wet etch and a final glancing angle Au layer is evaporated on the backside of the chip acting as a ground plane. In step \textbf{(p)}, the front-side and back-side gold layers are removed individually for improved etch uniformity by temporarily bonding the chip to a Si carrier. A microphotograph after the backside Au etch in \textbf{(p)} is provided.}
        \label{Fig_Fab}
        % \vspace*{-0.1in}
    \end{figure}
    
    To demonstrate the lithographic bandwidth adaptability, filter layout variations are fabricated with shunt resonator YIG widths of $\SI{12}{\micro\meter}$, $\SI{14}{\micro\meter}$, $\SI{16}{\micro\meter}$, and $\SI{18}{\micro\meter}$, each producing distinct demagnetizing fields and spinwave wave vectors. The small signal SW ladder filter responses for the devices in Fig. $\ref{Fig_Design}$e-g are measured using the experimental setup and settings described in Supplementary Information \ref{sup-Measurement_Setup}. An electromagnet provides an adjustable, uniform, out-of-plane magnetic bias for each device. Fig. $\ref{Fig_filterResponse}$ shows the measured performance for the fabricated 3\textsuperscript{rd}-order filters and Fig. $\ref{Fig_filterResponse}$a highlights the filter response for device D4 (Fig. $\ref{Fig_Design}$e) with a $\SI{12}{\micro\meter}$ shunt resonator YIG width near the passband as the external magnetic field is swept from $\SI{373.4}{\milli\tesla}$ to $\SI{904.2}{\milli\tesla}$. As discussed in detail in Supplementary Information \ref{sup-resonator_design_modeling}, the SW resonators show a frequency dependent $Q$-factor with an increasing trend over the magnetic bias. Therefore the filter insertion loss increases as the center frequency is tuned towards low frequency as evident in Fig. $\ref{Fig_filterResponse}$b. The filter's $\SI{3}{\decibel}$ bandwidth in Fig. $\ref{Fig_filterResponse}$c remains nearly constant as expected above $\SI{14}{\giga\hertz}$ for device D4. At low frequencies, the resonator coupling is no longer sufficiently high to support the large fractional bandwidth of the filter so the passband ripple increases and the bandwidth degrades. From Fig. $\ref{Fig_filterResponse}$c, the filters with a smaller bandwidth have an extended lower frequency tuning range due to their reduced resonator coupling requirements. The filter rejection in Fig. $\ref{Fig_filterResponse}$d is calculated as the worst case $S_{21}$ at $5$ bandwidths away from the center frequency. The filters with wider YIG resonators show superior rejection because the shunt transducers have a lower impedance (due to the wider Au transducers), resulting in a stronger impedance contrast relative to the series resonator. The wide transducer also strongly suppress spurious SW modes \cite{devitt_distributed_2024} with the exception of one SW mode in the series resonator appearing at higher frequencies (Fig. S$\ref{sup-SI_measured_resonator_response}$b). As a result, the filter responses are free from spurious passbands with the exception of the one spur on the low-side of the passband which becomes evident above $\SI{18}{\giga\hertz}$. The filter rejection can be substantially improved by including additional resonator elements in the filter at the cost of insertion loss and bandwidth. Fig. $\ref{Fig_Design}$f shows the layout for a 5\textsuperscript{th}-order filter with  3-series and 2-shunt resonators and Supplementary Information \ref{sup-5th_order_filter} discusses its design and performance compared to the 3\textsuperscript{rd}-order filters.\\

    The linearity and power handling capability for the SW ladder filters are measured using device D4 in Supplementary Information \ref{sup-Si_sec_nonlinearity}. The filters show a high input referred 3\textsuperscript{rd}-order intercept point (IIP3) compared to state-of-the-art magnetostatic surface wave (MSSW) YIG filters in Table \ref{Performance_Comparison} with an in-band IIP3 of $\geq 11.0 \text{ dBm}$ at $\SI{716.3}{\milli\tesla}$, $\geq 9.6 \text{ dBm}$ at $\SI{745.1}{\milli\tesla}$, $\geq 10.2 \text{ dBm}$ at $\SI{785.3}{\milli\tesla}$ using a $\SI{1}{\mega\hertz}$ tone spacing. Furthermore, the filter shows no power dependent performance or compression within the passband up to an input power of $5 \text{ dBm}$ as highlighted in Fig. S$\ref{sup-SI_Fig_PowerHandling}$ and Fig. S$\ref{sup-S12_PowerSweep_IL_BW}$.\\

    \begin{figure}[!b]
        \centering
        \includegraphics[width=\textwidth]{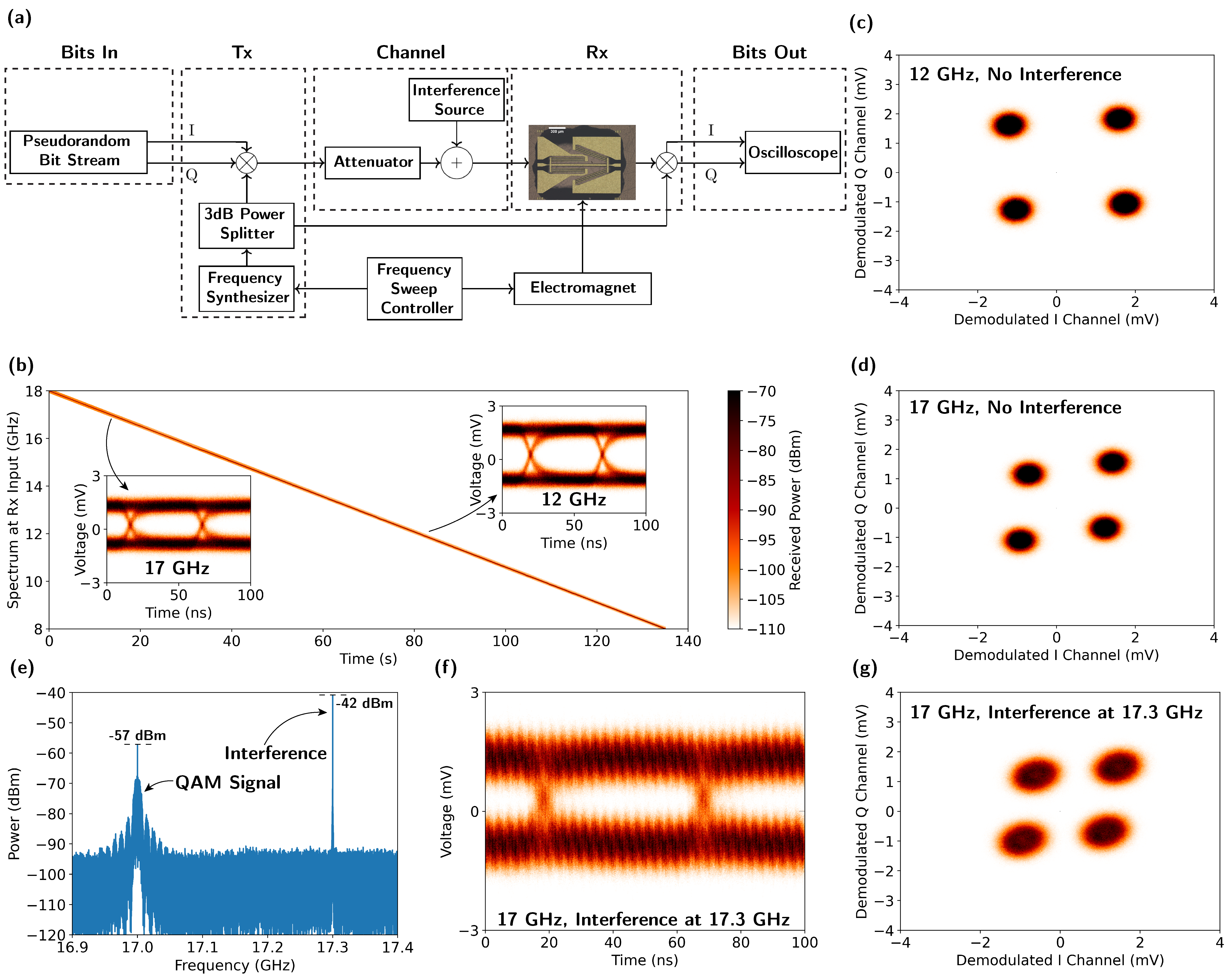}
        \caption{\textbf{Frequency agile radio.} \textbf{(a)} Functional block diagram of the frequency tunable radio measurement setup. Using QAM modulation, a $20 \text{ Mbps}$ pseudorandom bit steam is upconverted to $f_0$ using an IQ mixer. The QAM modulated RF signal is passed through a $\SI{20}{\decibel}$ attenuator acting as an AWGN channel and an interference source is added. The received signal passes through a 3\textsuperscript{rd}-order tunable YIG filter (Fig. \ref{Fig_Design}e) before being demodulated down to base band using a second IQ mixer. The transmit and receive IQ mixers share the same LO at $f_0$. A frequency sweep controller continuously tunes $f_0$ and the bias of an electromagnet such that the YIG filter's passband is always aligned with the communication channel. \textbf{(b)} Plot of the measured received frequency spectrum over time at the input to the YIG filter (through a \SI{10}{\decibel} coupler) tuning from $\SI{18}{\giga\hertz}$ to $\SI{8}{\giga\hertz}$ in $\SI{0.5}{\mega\hertz}$ steps. Eye diagrams of the demodulated I channel data streams are shown at $f_0=\SI{12}{\giga\hertz}$ and $f_0=\SI{17}{\giga\hertz}$. \textbf{(c)-(d)} show the demodulated IQ constellation diagrams at \textbf{(c)} $f_0=\SI{12}{\giga\hertz}$ and \textbf{(d)} $f_0=\SI{17}{\giga\hertz}$. An amplitude modulated gaussian noise interference signal at $f_i$ is added to the transmitted QAM-modulated data and \textbf{(e)} shows the power spectrum at the input to the filter with $f_0=\SI{17}{\giga\hertz}$ and $f_i=\SI{17.3}{\giga\hertz}$.  \textbf{(f)-(g)} Plot the demodulated Q channel eye diagram \textbf{(f)} and IQ constellation diagram \textbf{(g)} after adding the out-of-band interference. Videos of the tunable radio are available in the supplementary material and additional information about this experiment is provided in Supplementary Information \ref{sup-Si_sec_radio}. }
        \label{Fig_radio}
        % \vspace*{-0.1in}
    \end{figure}
    
    The fabrication process presented in Fig. $\ref{Fig_Fab}$ is critical for achieving both the resonator frequency separation and high resonator $k_{eff}^2$ required to implement a ladder filter using spinwaves. The filters, resonators, and de-embedding structures are fabricated on a $\SI{3}{\micro\meter}$-thick YIG-on-GGG chip measuring $\SI{15}{\milli\meter}\times \SI{15}{\milli\meter} \times \SI{537}{\micro\meter}$. The front-side/back-side micromachining fabrication is based on the hairclip YIG resonator process introduced in \cite{tiwari_high-performance_2025}. The ladder filter designs from Fig. $\ref{Fig_Design}$ do not require thru-GGG vias which greatly simplifies the fabrication resulting in high yield and consistent device performance across the chip. The Au transducer metal quality and roughness are enhanced by fine tuning the Ti hard mask thicknesses and wet etch timings for the thin shunt YIG fin resonators. Improvements to the fabrication uniformity also come from an optimized wax bonding and thinning process in step $\textbf{(l)}$. A micrograph of the fabricated chip is provided in Fig. $\ref{Fig_Design}$ along with micrographs of the fabricated 3\textsuperscript{rd}-order 1-series, 2-shunt and 5\textsuperscript{th}-order 3-series, 2-shunt filter layouts.
    
\section*{Tunable Radio Demonstration}

The wide frequency tunability of the SW ladder filter makes them attractive for frequency agile radio systems where the operating frequency dynamically adjusts in response to interference or continuously shifts such as in frequency hopped spread spectrum (FHSS) systems \cite{torrieri_principles_2018}. In congested spectral environments or scenarios where interference is anticipated, continuously frequency hopping provides significant enhancement to the bit error rates (BER) based on the system's processing gain, $\eta$, defined as the ratio of the hopping bandwidth to channel bandwidth \cite{torrieri_principles_2018, torrieri_frequency-hopping_2003, tse_fundamentals_2005, proakis_fundamentals_2005, holmes_spread_2007}. Alternatively, a frequency agile radio can sense and minimize out-of-band interference signals by dynamically adjusting its operating frequency. In both cases, a tunable bandpass filter can be integrated between the receiver's antenna and first amplification stage, providing a direct improvement to the signal to noise and interference ratio (SNIR) by attenuating any out-of-band jamming/interference without sacrificing tuning range.

To demonstrate the SW ladder filter's use in such a system, a simplified frequency agile radio is constructed in Fig. $\ref{Fig_radio}$a. A quadrature amplitude modulated (QAM) pseudorandom bit stream at $20 \text{ Mbps}$ is transmitted over an additive white Gaussian noise (AWGN) channel (implemented as a $\SI{20}{\decibel}$ attenuator). At the radio's receiver, the noisy signal is filtered using the tunable 3\textsuperscript{rd}-order SW filter and demodulated into separate in-phase (I) and quadrature (Q) data streams. The local oscillator (LO) frequency, $f_0$, constantly varies over time and an electromagnet tunes the YIG filter such that the passband is always aligned to the communication channel. The demodulated eye diagrams and IQ constellations are shown in Fig. $\ref{Fig_radio}$b-d. In this experimental setup, the frequency tuning speed is primarily limited by the electromagnet's inductance and not the SW ladder filter. With a fast-tuning magnet bias, this YIG filter could be integrated into the receiver for fast hopping frequency agile radio to provide a strong enhancement to the SNIR. 

The performance of the YIG filter is also evaluated when an out-of-band interference signal is present. A strong amplitude modulated gaussian noise interference tone is combined with the transmitted QAM-data as shown in Fig. $\ref{Fig_radio}$a. The received power spectrum in Fig. $\ref{Fig_radio}$e is monitored at the input to the filter using a $\SI{10}{\decibel}$ coupler and spectrum analyzer. The plot illustrates the relative signal and interference levels at the input of the receiver. Fig $\ref{Fig_radio}$f and Fig $\ref{Fig_radio}$g show the demodulated Q channel eye diagram and IQ constellation respectively with the out-of-band interference $\SI{300}{\mega\hertz}$ away from the channel in the filter's stopband. Despite the strong out-of-band interference, the YIG filter provides sufficient attenuation to preserve high SNIR for the signal of interest. Further discussion about the radio demonstration in Fig. $\ref{Fig_radio}$ is available in Supplementary Information \ref{sup-Si_sec_radio}. Videos of the radio's operation are also included in the supplementary materials. 
 
\section*{Discussion}
% \section*{Conclusion}
    
SW ladder filters show strong potential for integration in frequency agile radios and commercial 5G and 6G systems due to their superior performance in the FR3 frequency range $\SI{7.125}{\giga\hertz}$ -- $\SI{24.25}{\giga\hertz}$. Their miniaturized footprint, high power handling, inherent tunability, and large bandwidths make them ideal to replace arrays of fixed-frequency filters required to cover a number of disjoint frequency bands. The key enabling technology for these filters is the YIG micromachining and deep GGG etching fabrication process optimized in this work. This process shows high device yield and wafer-scale process control for uniformity. By precisely patterning the YIG films, the component SW resonators require only one external magnetic bias resulting improved performance while simultaneously reducing packaging volume, complexity, and cost. Recent developments in micromagnetic packaging techniques including \cite{wang_temperature_2024, du_frequency_2024} provide a roadmap towards fully integrated tunable SW filters. For a frequency agile radio application, we demonstrate the SW ladder filters in a continuously-tuned radio receiver. The filters show excellent demodulated bit streams over an octave of center frequency tuning and exhibit strong immunity to interference in an adjacent channel.
 
\section*{Methods}

\subsection*{Device Fabrication}
A detailed fabrication process flow is provided in Fig. $\ref{Fig_Fab}$. Details regarding the YIG ion milling steps in \textbf{(a)}-\textbf{(c)} are available in \cite{dai_octave-tunable_2020, devitt_edge-coupled_2024, feng_micromachined_2023}. A discussion on the patterned electroplating process in steps \textbf{(d)}-\textbf{(j)} is provided in \cite{devitt_distributed_2024}. The deep anisotropic GGG etch used in step \textbf{(o)} is accessible in \cite{tiwari_high-performance_2025}.

\subsection*{Filter and Resonator Measurements}
Supplementary Information \ref{sup-Measurement_Setup} discusses the experimental setup and measurement setting in detail for all small signal S-parameter data and Supplementary Information \ref{sup-Si_sec_nonlinearity} details the power handling and nonlinearity measurements. Supplementary Information \ref{sup-Si_sec_radio} discusses the experimental setup for the frequency agile radio.

\section*{Data Availability}
	% The raw S-parameter and non-linearity data generated in this study have been deposited in the Zenodo database \\url{ALSDKJFHA}. Any other data is available upon request.
    All raw data generated in this study will be released in a Zenodo repository upon publication.

    \balance
    
\section*{Acknowledgments}
    Chip fabrication was performed at the Birck Nanotechnology Center at Purdue. Filter and resonator  measurements were performed at Seng-Liang Wang Hall at Purdue. \\

    The Purdue authors would like to thank  Dr. Bryan Bosworth, Dr. Nick Jungwirth, and Dr. Nate Orloff at NIST for their helpful discussions on filter nonlinearity and high-power S-parameter measurements. The authors also thank Dr. Todd Bauer and Dr. Zachary Fishman for helpful discussions.\\
    
    This research was developed with funding from the Air Force Research Laboratory (AFRL) and the Defense Advanced Research Projects Agency (DARPA) under the COFFEE program. A portion of the Purdue research is also supported by funding from the Army Research Laboratory (ARL) and Keysight Technologies. The views, opinions and/or findings expressed are those of the authors and should not be interpreted as representing the official views or policies of the Department of Defense or the U.S. Government. This manuscript is approved for public release; distribution A: distribution unlimited.\\

\section*{Author Contributions}

C.D., S.T., and B.Z. contributed to the filter concept and fabrication process development. C.D. simulated and measured the resonators and filters. Manuscript was prepared by C.D. with input from S.T., B.Z., and S.A.B.

\section*{Competing Interests}
    The authors declare no competing interests.

	% \vfill

 \bibliographystyle{IEEEtran}
\bibliography{ARROW_bib}

\end{document}

\bstctlcite{IEEEexample:BSTcontrol}

    \title{Spinwave Bandpass Filters for 6G Communication: Supplementary Information}
	
	\author{Connor Devitt$^{1\ast}$ \orcidlink{0000-0002-3394-6842}, Sudhanshu Tiwari$^{1}$ \orcidlink{0000-0002-5499-0877}, Bill Zivasatienraj$^{2}$ \orcidlink{https://orcid.org/0000-0002-0522-4869}, Sunil A. Bhave$^{1\ast}$ \orcidlink{0000-0001-7193-2241}
		% <-this % stops a space
            \\ $^{1}$OxideMEMS Lab, Elmore Family School of Electrical and Computer Engineering, Purdue University, West Lafayette, IN 47907 USA. 
            % \\ $^{2}$Amazon Web Services, Pasadena, CA 91106.
            \\ $^{2}$FAST Labs\textsuperscript{TM}, BAE Systems, Inc., Nashua, NH 03060 USA. 
            \\ $^\ast$e-mail: devitt@purdue.edu, bhave@purdue.edu
    }

    \maketitle

    \tableofcontents
    
    \clearpage

\section{Supplementary Information: Spinwave and Acoustic Wave Resonator Modeling}
    \label{Acoustic_analogy}

    \begin{figure}[!b]
        \centering
        \vspace*{-0.1in}
        \includegraphics[width=\textwidth]{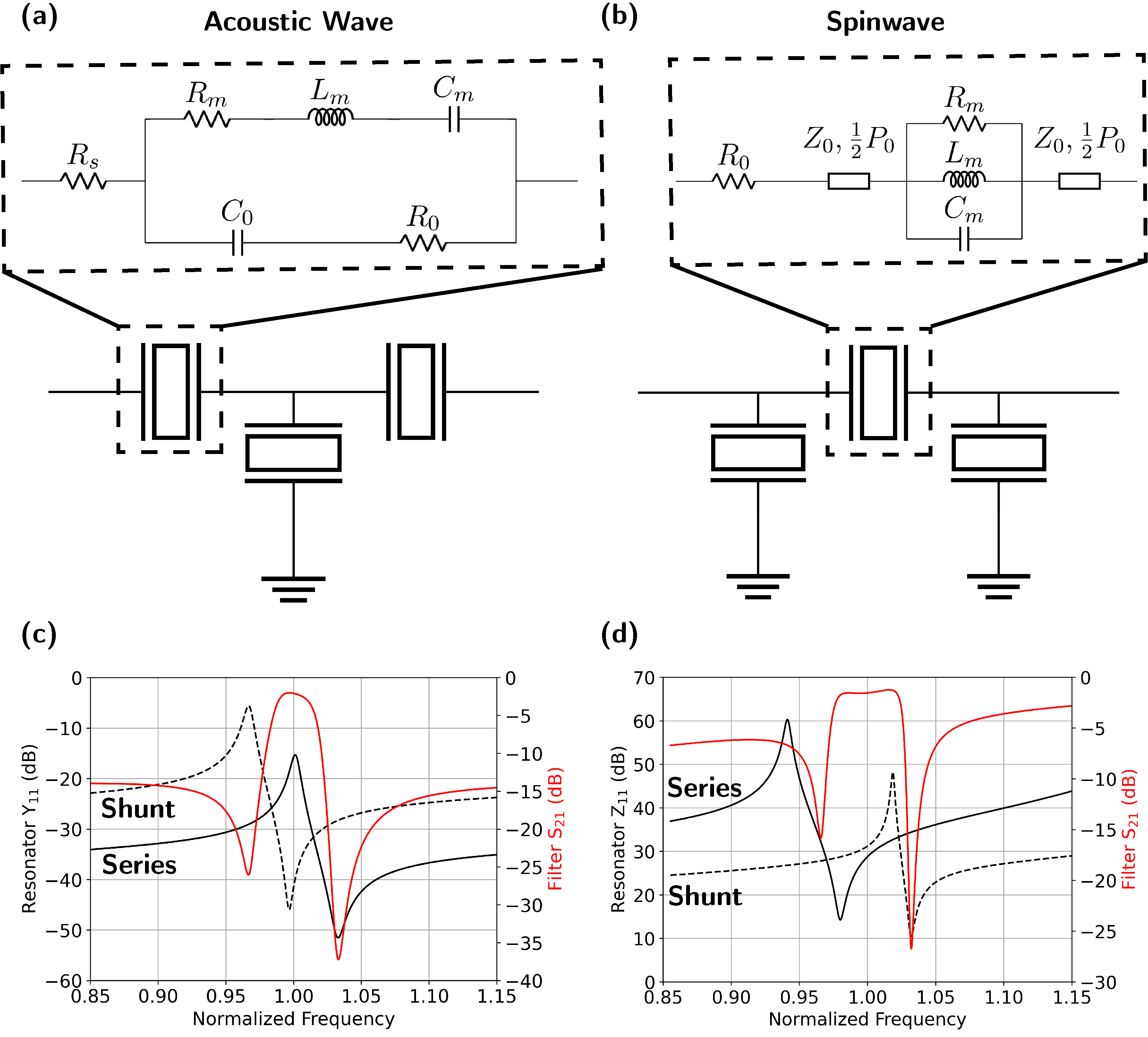}
        \caption{\textbf{Modeling comparison between acoustic wave and spinwave resonators and ladder filters} \textbf{(a)} Acoustic wave and \textbf{(b)} spinwave 3-resonator ladder filter schematics where each component resonator is modeled using \textbf{(a)} the mBVD model \cite{larson_modified_2000} and \textbf{(b)} 2-port distributed SW model. Series and shunt \textbf{(c)} acoustic resonator admittance and \textbf{(d)} spinwave resonator impedance response along with the corresponding 3-resonator filter response. }
        \label{SI_Fig_Acoustic_SW_compare}
        \vspace*{-0.1in}
    \end{figure}
    
     Fig. S$\ref{SI_Fig_Acoustic_SW_compare}$ illustrates the modeling differences between acoustic wave and spinwave resonators and filters. In Fig. S$\ref{SI_Fig_Acoustic_SW_compare}$a, the acoustic resonators use a capacitive transducer to excite a particular acoustic mode and the mBVD model captures both the electrical response in the form of a series resistance ($R_s$) and parallel capacitance ($C_0$) as well as the mechanical response in the form of a motional series resonator network ($R_m$, $C_m$, $L_m$). Far away from the acoustic resonance, the acoustic ladder filter behaves as a capacitive divider network due to the alternating series $C_{0,s}$ and shunt $C_{0,p}$. Therefore, the out-of-band rejection is determined by the impedance contrast of $C_{0,s}$ and $C_{0,p}$ as well as the number of resonator stages in the filter. The resonators exhibit a high admittance series resonance at $f_s$ corresponding to the motional branch $R_m$, $C_m$, and $L_m$ and a low admittance parallel anti-resonance at $f_p$ due to $C_0$ in parallel with the motional branch. The filter passband is formed by frequency shifting the shunt resonator such that $f_s$ of the series resonator aligns with $f_p$ of the shunt resonator giving a low impedance path from port 1 to port 2 and a high impedance path to ground \cite{hashimoto_rf_2009}.

     In the case of a spinwave resonators, the response is described using the 2-port distributed SW model in Fig. S$\ref{SI_Fig_Acoustic_SW_compare}$b. The inductive transducers are modeled by the series $R_0$ with a transmission line with impedance $Z_0$ and a total physical length $P_0$ while the spinwave resonance is described using the parallel $R_m$, $C_m$, and $L_m$. Similar to the mBVD model, this spinwave model shows both a high impedance resonance and low impedance anti-resonance. The filter passband is formed by introducing a frequency shift between the series and shunt resonators in the way as acoustic ladder filters. Fig. S$\ref{SI_Fig_Acoustic_SW_compare}$c-d illustrate how the alternating series and shunt resonators give rise to a bandpass filter response by plotting the input admittance and impedance of the component resonators overlaid with a 3-resonator filter response.

    \section{Supplementary Information: Spinwave Dispersion}
    \label{spinwave_dispersion}

\begin{figure}[!h]
        \centering
        % \vspace*{-0.1in}
        \includegraphics[width=0.75\textwidth]{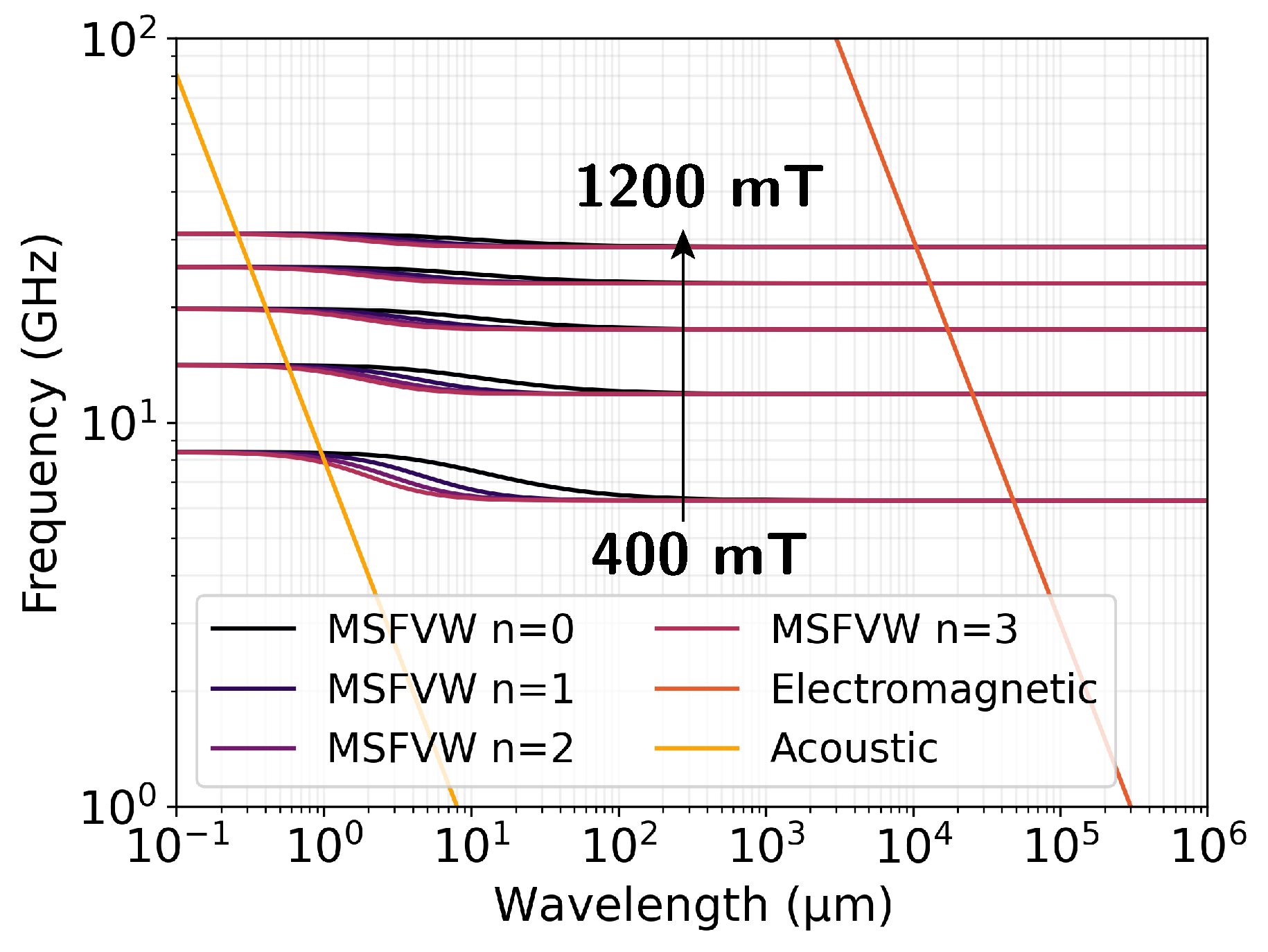}
        \caption{\textbf{Dispersion of magnetostatic forward volume waves.} Dispersion of MSFVW modes with a film thickness of $t=\SI{3}{\micro\meter}$ as compared to electromagnetic waves with a group velocity of $3\times 10^8\text{ m/s}$ and acoustic waves with group velocity of $\SI{8000}{\meter/\second}$. }
        \label{SI_Fig_MSFVW_dispersion}
        % \vspace*{-0.1in}
    \end{figure}

    The dispersion for magnetostatic forward volume waves (MSFVW) from \cite{stancil_theory_1993} is given as
    \begin{align}
        \tan\left(\frac{k_{mn}t}{2}\sqrt{-\left(1 + \chi\right)} - \frac{n\pi}{2}\right) &= \frac{1}{\sqrt{-\left(1 + \chi \right)}},
        \label{eqn_MSFVW_general_dispersion}
    \end{align}
    where $k_{mn}$ is the SW wavevector, $t$ is the film thickness, $n$ is the mode number and $\chi$ is defined as
    \begin{align}
        \chi &= \frac{\omega_0\omega_m}{\omega_0^2 - \omega^2},
    \end{align}
    with $\omega_0 = \gamma_m\mu_0 H_{DC}^{eff}$, $\omega_m=\gamma_m\mu_0 M_s$, and $\omega$ is the wave frequency. The dispersion relation for the first few MSFVW modes (ignoring exchange) are plotted in Fig. S$\ref{SI_Fig_MSFVW_dispersion}$ alongside an electromagnetic wave with group velocity $\nu_g=3\times 10^8\text{ m/s}$ and a non-dispersive acoustic wave with $\nu_g=\SI{8000}{\meter/\second}$.

\section{Supplementary Information: Measurement Setup}
    \label{Measurement_Setup}

    \begin{figure}[!b]
        \centering
        \vspace*{-0.1in}
        \includegraphics[width=\textwidth]{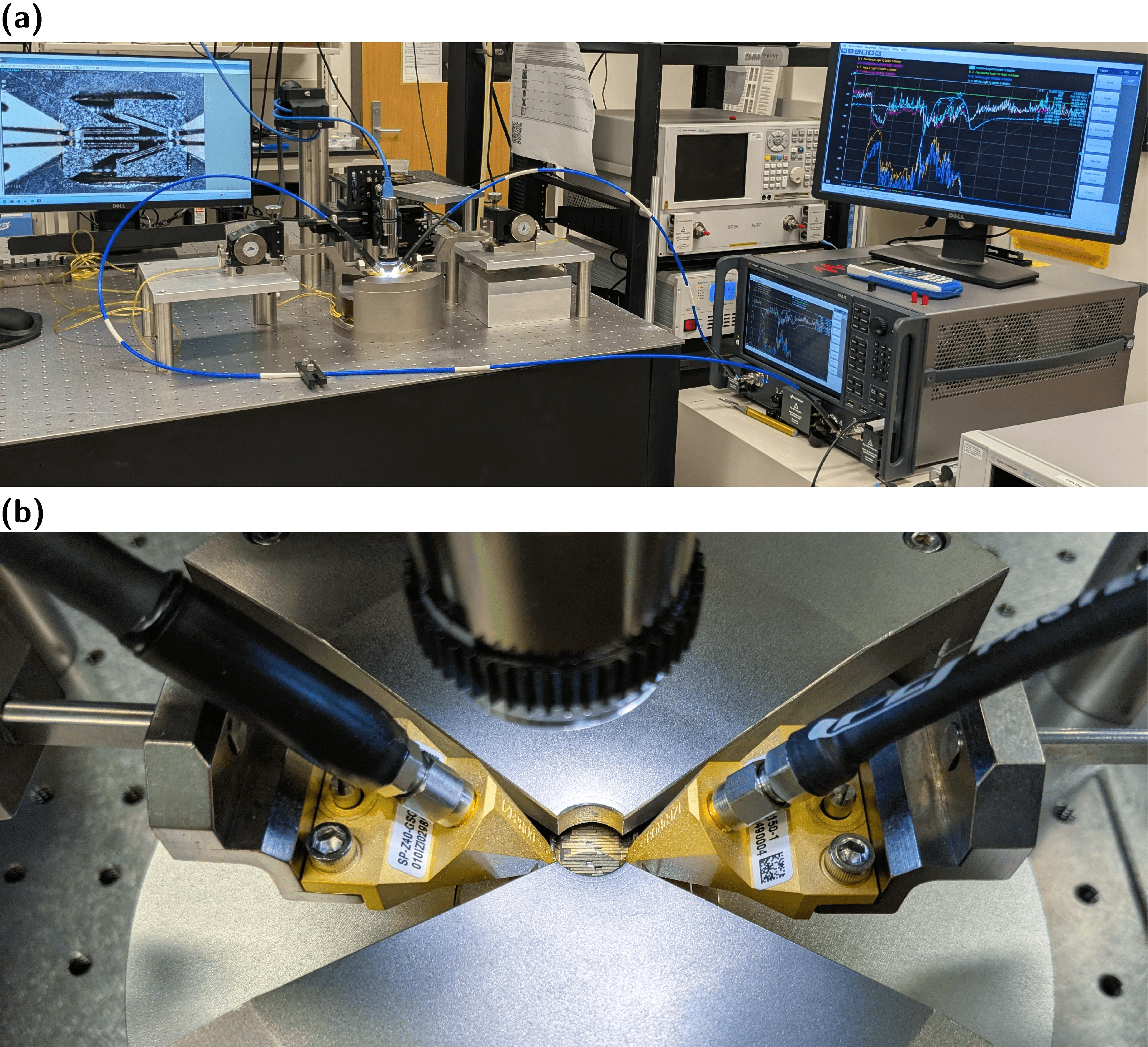}
        \caption{\textbf{SW ladder filter experimental setup.} \textbf{(a)} Photograph of the full experimental setup. A PNA-X (N5244B) is connected to two GSG non-magnetic RF probes through \SI{2.92}{\milli\meter} cables.
         A digital objective is used to align and land the RF probes. A current source controls the electromagnet's out-of-plane bias field and a temperature probe continuously monitors the coil's temperature rise. \textbf{(b)} Shows a zoomed in view of the filter chip resting on the electromagnet's pole. }
        \label{SI_Fig_Measurement_setup}
        \vspace*{-0.1in}
    \end{figure}
    
    Fig. S$\ref{SI_Fig_Measurement_setup}$ shows the physical measurement setup for both the filter s-parameter and the intermodulation distortion measurements. The 3\textsuperscript{rd}-order spinwave (SW) filter and resonator measurements use non-magnetic $\SI{150}{\micro\meter}$ GSG RF probes while the 5\textsuperscript{th}-order filter measurements use $\SI{150}{\micro\meter}$ GS probes. All s-parameter data is measured with an input power of $-20 \text{ dBm}$ and an intermediate frequency bandwidth (IFBW) of $\SI{5}{\kilo\hertz}$ unless otherwise noted. A short-open-load-thru (SOLT) calibration on a separate calibration substrate is performed to move the measurement reference plane to the probe tips. Details regarding the intermodulation distortion measurement settings are provided in  Supplementary Information \ref{Si_sec_nonlinearity}. The filter chip is bonded to a Si carrier chip using a Cyrstalbond adhesive for mechanical support. Prior to each measurement, the device under test (DUT) is aligned to the center of the electromagnet's $\SI{10}{\milli\meter}$ diameter pole face to ensure the YIG films are uniformly biased. The Si carrier is attached the the electromagnet using double-sided tape. The probe pads are de-embedded from the 3\textsuperscript{rd}-order SW filter and resonator measurements using device D5 consisting of two back-to-back GSG probe landings identical to the geometry used in each filter and resonator layout.  

    \begin{figure}[!h]
        \centering
        \vspace*{-0.1in}
        \includegraphics[width=0.75\textwidth]{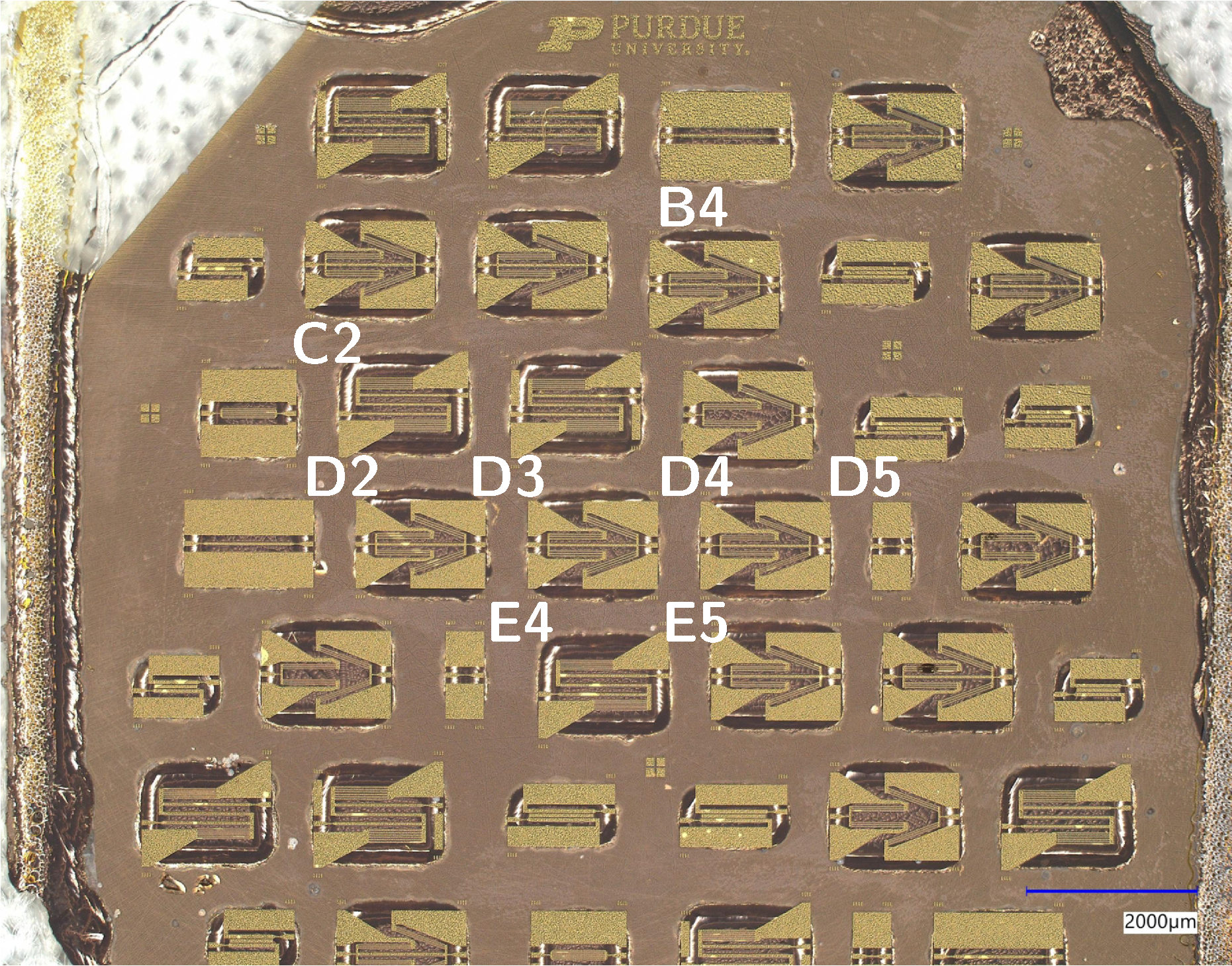}
        \caption{\textbf{SW ladder filter chip layout.} Chip micrograph labeling the de-embedding structure (D5), 3\textsuperscript{rd}-order filters (D2, D3, D4, E5, and B4), and 5\textsuperscript{th}-order filters (C2 and E4) used throughout this work.}
        \label{SI_Fig_device_labels}
        \vspace*{-0.1in}
    \end{figure}

    % \clearpage

\section{Supplementary Information: Resonator Design and Modeling}
    \label{resonator_design_modeling}

\subsection{Resonator Coupling and Q-factor}
    The effective resonator coupling ($k_{eff}^2$)  and $Q$-factors reported in this work use the definitions 
    \begin{equation}
        k_{eff}^2 = \frac{\pi}{2}\frac{f_p}{f_s}\cot\left(\frac{\pi}{2}\frac{f_p}{f_s}\right),
        \label{coupling_equation}
    \end{equation}
    and 
    \begin{equation}
        Q = \frac{f_p}{\Delta f},
    \end{equation}
    where $f_p$ is the SW resonance frequency, $f_s$ is the anti-resonance frequency, and $\Delta f$ is the $\SI{3}{\decibel}$ bandwidth of the resonance at $f_p$ from the resonator's impedance spectrum ($Z_{11}$). These definitions are consistent with previously reported metrics \cite{tiwari_high-performance_2025, devitt_distributed_2024, devitt_edge-coupled_2024}.

\subsection{Resonator Tuning}
\begin{figure}[!b]
        \centering
        \includegraphics[width=\textwidth]{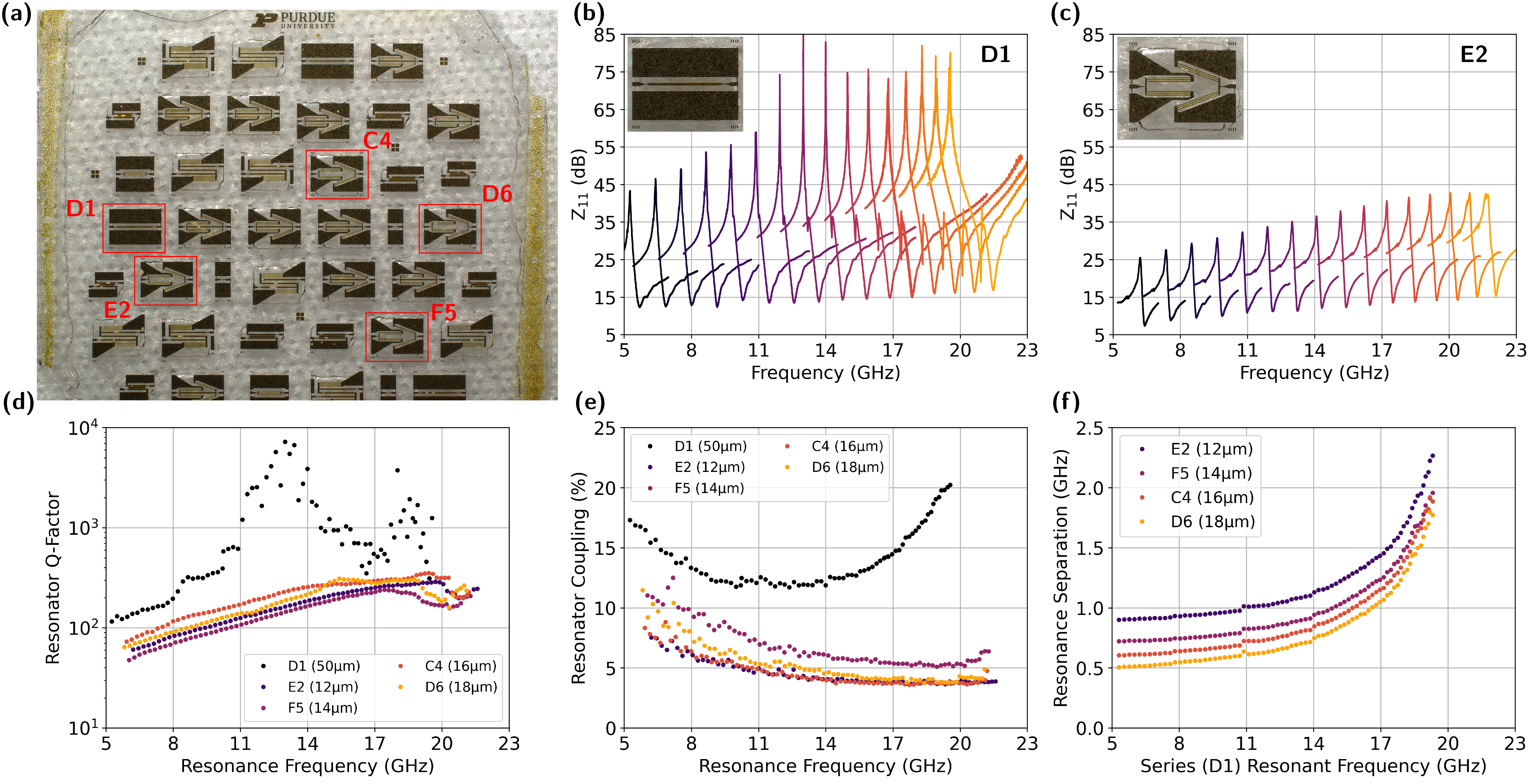}
        \caption{\textbf{Measured frequency response and performance of series and shunt SW hairclip resonators.} \textbf{(a)} Chip microphotograph prior to backside gold deposition showing the individual series resonator (D1) and shunt resonators (C4, D6, E2, and F5). Each shunt resonator layout is identical to the 3\textsuperscript{rd}-order SW ladder filter layout from Fig. $\ref{main-Fig_Design}$e, but the series resonator is omitted. \textbf{(b)} Input impedance of the two-port series resonator D1 (with a YIG width of $\SI{50}{\micro\meter}$) over magnetic bias after terminating port 2 with an ideal short and de-embedding the probe pads using device D5. \textbf{(c)} Input impedance at port 1 of the shunt resonator E2 (with a YIG width of $\SI{12}{\micro\meter}$) over magnetic bias with de-embedded probe pads. \textbf{(d)} Measured $\SI{3}{\decibel}$ $Q$-factor at $f_p$ and \textbf{(e)} effective coupling coefficient, $k_{eff}^2$, using Eq. $\ref{coupling_equation}$ for each SW hairclip resonator over magnetic bias. To extract the high $Q$-factors of device D1, the measurement frequency range was segmented into three and swept using $40,001$ points. Additionally a Savitzky-Golay filter with a $\SI{5}{\mega\hertz}$ window is applied in post-processing to reduce the measurement noise around the resonance bandwidth in the $Z_{11}$ spectrum.  \textbf{(f)} Measured resonance frequency separation over shunt resonator width and magnetic field relative to the resonance of D1.}
        \label{SI_measured_resonator_response}
        % \vspace*{-0.1in}
    \end{figure}

\begin{figure}[!t]
        \centering
        \includegraphics[width=0.75\textwidth]{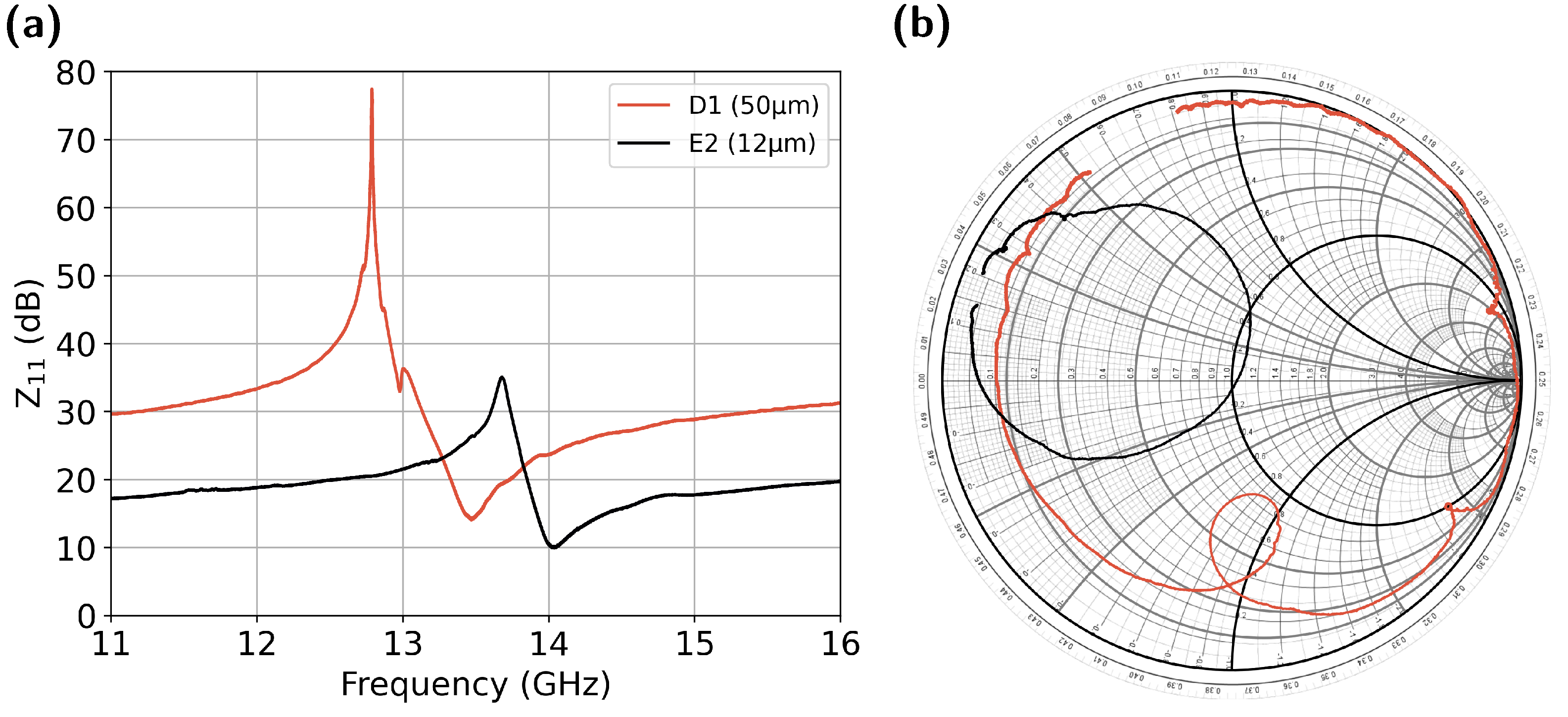}
        \caption{\textbf{Measured SW hairclip resonator impedance response.} \textbf{(a)} Resonator frequency response of device D1 and E5 biased at $\SI{632.4}{\milli\tesla}$. \textbf{(b)} Smith chart plot of two resonator's $S_{11}$ at the same bias. Similar to Fig. S$\ref{SI_measured_resonator_response}$ the probe pads are de-embedded from both devices and port 2 of D1 is terminated with an ideal short. At $\SI{632.4}{\milli\tesla}$, device D1 shows measured $Q=2664$, $k_{eff}^2=11.88\%$, and $f_0=\SI{12.79}{\giga\hertz}$ while  device E2 with a parallel array of 6 YIG fins shows measured $Q=159$, $k_{eff}^2=6.17\%$, and $f_0=\SI{13.68}{\giga\hertz}$.}
        \label{SI_Fig_Resonator_Smith}
        % \vspace*{-0.1in}
    \end{figure}

    To predict the SW ladder filter performance over a wide range of magnetic biases, the distributed SW resonator model in Fig. $\ref{main-Fig_Design}$b must be tuned to capture the frequency and magnetic field dependence of the resonator's effective coupling and $Q$-factor. The ferromagnetic resonance (FMR) linewidth ($\Delta H$) is given by \cite{beaujour_ferromagnetic_2009, kittel_introduction_nodate, stancil_theory_1993} as
    \begin{equation}
        \Delta H = \Delta H_0 + \frac{4\pi}{\gamma_m\mu_0}\alpha f,
    \end{equation}
    where $\Delta H_0$ is the inhomogeneous broadening, $\gamma_m\mu_0 /2\pi=28\si{\tfrac{\mega\hertz}{\milli\tesla}}$ is the gyromagnetic ratio, $\alpha$ is the Gilbert damping, and $f$ is the resonance frequency. Multiplying through by $\gamma_m / 2\pi$ gives the resonance bandwidth as 
    \begin{equation}
        \Delta f = \Delta f_0 + 2\alpha f.
    \end{equation}
    For liquid phase epitaxy (LPE) single-crystal YIG films, the Gilbert damping is on the order of $\alpha \approx 10^{-4}$ \cite{chumak_advances_2022, dubs_low_2020}. Based on the measured resonator responses in Fig. S$\ref{SI_measured_resonator_response}$ and Fig. S$\ref{SI_Fig_Resonator_Smith}$, the minimum resonance bandwidth is $\SI{49.5}{\mega\hertz}$ over all shunt resonators and $\SI{1.8}{\mega\hertz}$ for the series resonator. Due to the relatively large resonance bandwidths for these devices, the influence of the additional frequency dependent broadening term arising from Gilbert damping can be neglected. Therefore, the YIG resonators can be modeled with a constant bandwidth over frequency as
    \begin{equation}
        \Delta f = \frac{f_0}{R_m}\sqrt{\frac{L_m}{C_m}} = \frac{f_0}{R_m}\kappa,
    \end{equation}
    where $\kappa=\sqrt{\frac{L_m}{C_m}}$ is treated as a constant. Given a set of distributed SW resonator model parameters fitted from either measurement or finite element simulation, that model can be magnetically tuned by adjusting the fitted $C_m$, $L_m$, and $R_m$ to $C_m'$, $L_m'$, and $R_m'$ using
    \begin{align}
        C_m' &= \frac{1}{2\pi\kappa f_0'},\\
        L_m' &= \kappa^2 C_m',\\
        R_m' &= \frac{\kappa f_0}{\Delta f},
    \end{align}
    where $f_0'=\frac{1}{2\pi\sqrt{L_m'C_m'}}$ is the new resonance frequency. This method shows strong agreement with finite element simulations (Ansys HFSS) of SW resonators tuned with an out-of-plane magnetic bias.\\

\subsection{Modeled Resonator Response over YIG Width}

\begin{figure}[!t]
        \centering
        \includegraphics[width=\textwidth]{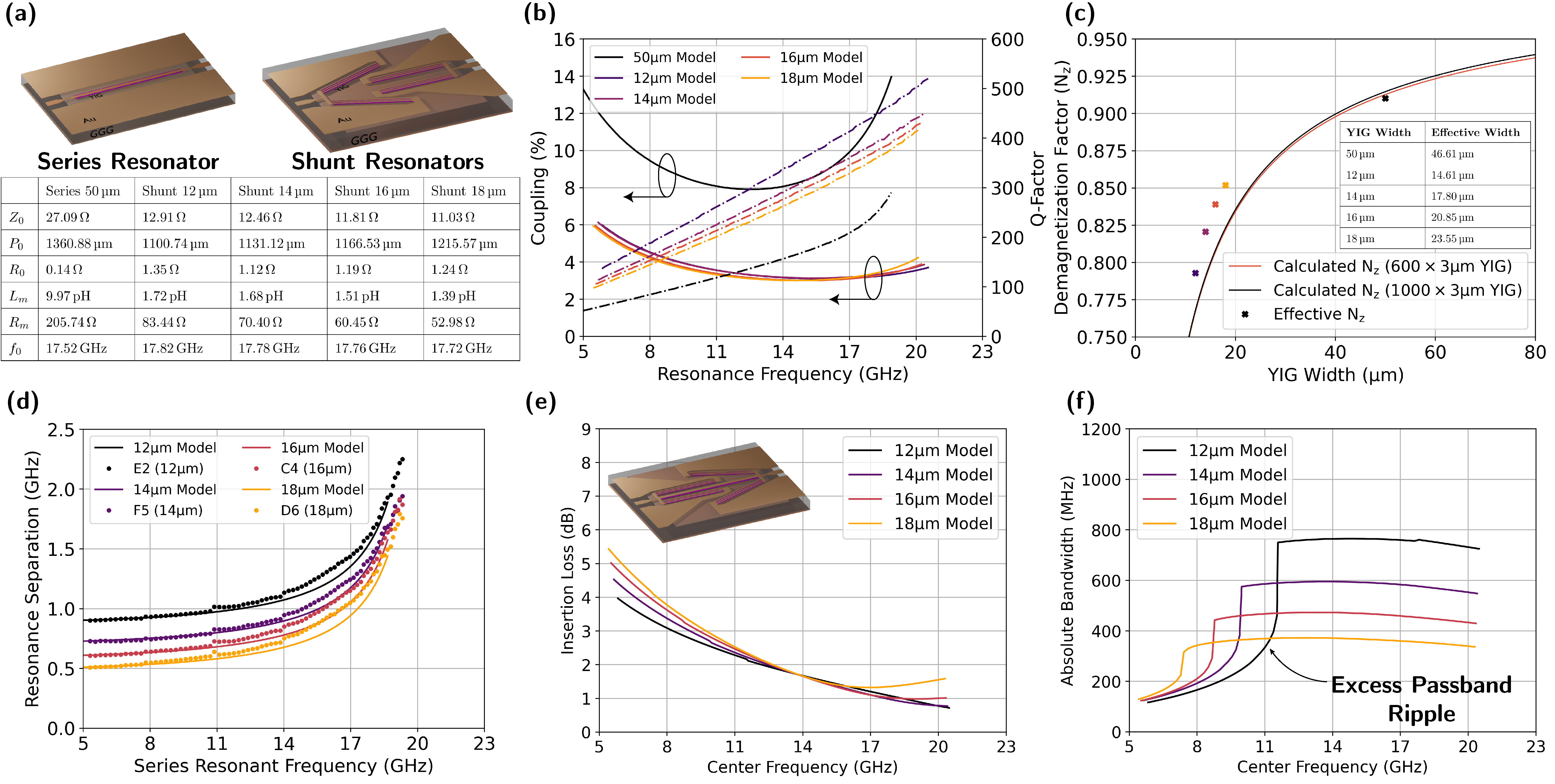}
        \caption{\textbf{Modeled SW hairclip resonator performance.} Each hairclip resonator geometry from Fig. S$\ref{SI_measured_resonator_response}$a is simulated using a finite element method at $\SI{792.3}{\milli\tesla}$ ignoring demagnetization effects and using a GGG membrane thickness of $\SI{10}{\micro\meter}$. \textbf{(a)} The simulated resonators are fitted to the distributed SW model from Fig. S\ref{SI_Fig_Acoustic_SW_compare}b and the extracted model parameters are listed for each YIG resonator width. \textbf{(b)} Extracted resonator coupling coefficients and $\SI{3}{\decibel}$ $Q$-factors over magnetic tuning for each fitted resonator model. \textbf{(c)} Analytic demagnetization factors ($N_z$) from \cite{aharoni_demagnetizing_1998} for the series and shunt resonators over YIG width compared to the effective $N_z$ fit from the measured resonator responses. \textbf{(d)} Comparison of the resonance frequency separation over YIG width (relative to the series resonator with $\SI{50}{\micro\meter}$ YIG width) between the measured resonator response and the fitted resonator models after applying the fitted $N_z$ factors. The resonator models are combined into a 3\textsuperscript{rd}-order 1-series, 2-shunt ladder filter using a port impedance of $Z_0=\SI{15}{\ohm}$ and the resulting filter insertion loss \textbf{(e)} and $\textbf{(f)}$ bandwidth are calculated over magnetic field tuning.}
        \label{SI_Fig_Demag}
        % \vspace*{-0.1in}
    \end{figure}

    To study the effective coupling, $Q$-factor, and resonance frequency dependence of the SW hairclip resonators on the width of the YIG mesa, 3D resonator models (pictured in Fig. S$\ref{SI_Fig_Demag}a$) for each fabricated resonator geometry are simulated through finite element method (Ansys HFSS) using a uniform out-of-plane magnetic and a GGG membrane thickness of $\SI{10}{\micro\meter}$. A distributed SW resonator model is fit to each model. $R_0$, $Z_0$, and $P_0$ are fit from the simulation at $\SI{0}{\milli\tesla}$ while $L_m$, $C_m$, and $R_m$ are fit from the simulation at $\SI{792.3}{\milli\tesla}$. The fitted model parameters for each resonator are listed in Fig. S$\ref{SI_Fig_Demag}a$. The finite element simulation captures the dispersion of the magnetostatic forward volume waves (MSFVW), but not the strongly geometry dependent demagnetization field so a correction factor is applied to the resonance frequency using \cite{aharoni_demagnetizing_1998} by tuning the resonator model using the previously described method. Fig. S$\ref{SI_Fig_Demag}$b show the modeled resonator $k_{eff}^2$ and $Q$-factor over the applied magnetic field. The simulated effective coupling exhibits a similar trend as the measured coupling in Fig. S$\ref{SI_measured_resonator_response}$e, but under predicts both the series and shunt resonator coupling. The higher measured coupling is primarily attributed to a thinner GGG membrane in the fabricated devices and is further addressed in Supplementary Information \ref{SI_GGG_Thickness}. The model under predicts the measured series resonator $Q$-factor in Fig. S$\ref{SI_measured_resonator_response}$d, but the shunt resonator $Q$-factor shows strong agreement with the model up to $\SI{18}{\giga\hertz}$ where the frequency dependent attenuation of the transducers becomes significant. \\

    \begin{figure}[!t]
        \centering
        \includegraphics[width=\textwidth]{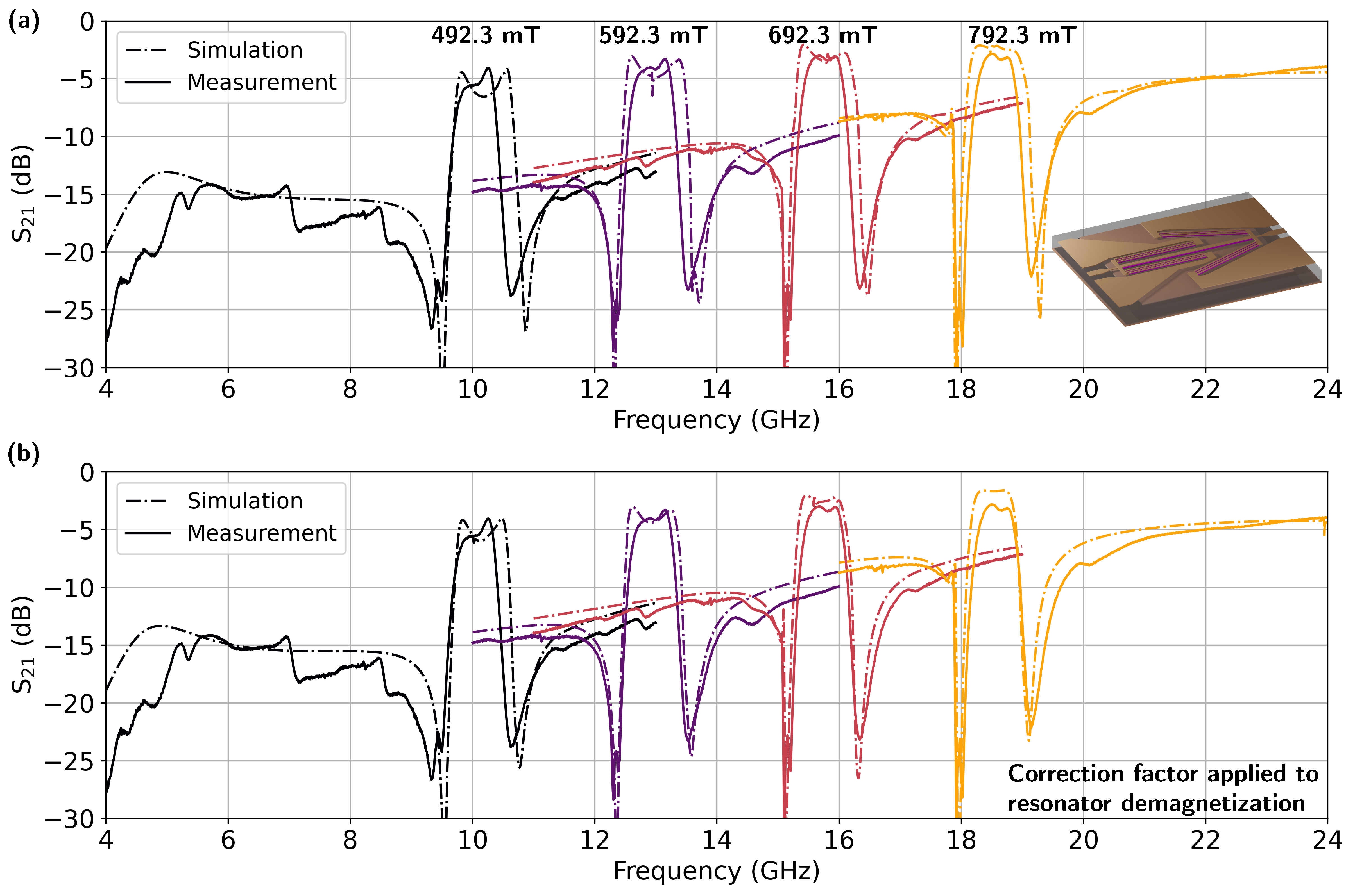}
        \caption{\textbf{Measured versus simulated SW ladder filter performance.} The SW ladder filter geometry of device D4 with a shunt resonator YIG width of $\SI{12}{\micro\meter}$ is simulated using a finite element method at $\SI{492.3}{\milli\tesla}$, $\SI{592.3}{\milli\tesla}$, $\SI{692.3}{\milli\tesla}$, and $\SI{792.3}{\milli\tesla}$ using a nominal GGG membrane thickness of $\SI{10}{\micro\meter}$. Demagnetization factor corrections are applied to each YIG mesa using \textbf{(a)} the analytic $N_z$ from \cite{aharoni_demagnetizing_1998} and using \textbf{(b)} the fitted $N_z$ factors from Fig. S$\ref{SI_Fig_Demag}$c. The probe pads are de-embedded and the port impedance is renormalized to $Z_0=\SI{15}{\ohm}$ for both the simulated and measured filter performance.}
        \label{SI_Fig_SimMeas_Filter}
        % \vspace*{-0.1in}
    \end{figure}
    
    Using the MSFVW dispersion from the finite element simulation and the demagnetization factor ($N_z$) calculated from the YIG mesa geometry using \cite{aharoni_demagnetizing_1998}, the distributed SW models predict a significantly higher resonance frequency separation between the series resonator and each shunt resonator as compared to the measured frequency separation in Fig. S$\ref{SI_measured_resonator_response}$f. A demagnetization field correction factor is optimized for each resonator by comparatively analyzing the resonance frequency separation over YIG resonator width of the series resonator relative to all shunt resonators as well as the shunt resonators relative to each other. Fig. S$\ref{SI_Fig_Demag}$d compares the measured frequency separation to the modeled resonance frequency separation over magnetic field with the optimized $N_z$. Fig. S$\ref{SI_Fig_Demag}$c illustrates the optimized $N_z$ for each resonator compared to the values calculated from \cite{aharoni_demagnetizing_1998} and computes the effective width of each resonator based on the optimized $N_z$. The series resonator shows close agreement with the calculated $N_z$, however, each shunt resonator has a much larger effective width leading to the smaller measured resonance frequency separation observed in Fig. S$\ref{SI_measured_resonator_response}$f. The demagnetization factor calculation provided in \cite{aharoni_demagnetizing_1998} is for a rectangular prism in isolation, however, the shunt resonators used in this work are grouped into arrays of 3--6 YIG fins separated by $\SI{10}{\micro\meter}$. Therefore, the larger $N_z$ and effective width for each shunt resonator is attributed to the close YIG fin spacing, but requires further study to determine the relationship between $N_z$ and physical resonator spacing.\\

    Fig. S$\ref{SI_Fig_Demag}$e and S$\ref{SI_Fig_Demag}$f present the modeled filter insertion loss (IL) and $\SI{3}{\decibel}$ bandwidth (BW) respectively for a 3\textsuperscript{rd}-order 1-series, 2-shunt SW ladder filter using the distributed resonator models from Fig. S$\ref{SI_Fig_Demag}$a and the optimized $N_z$ from Fig. S$\ref{SI_Fig_Demag}$c. Consistent with the measured filter performance in Fig. $\ref{main-Fig_filterResponse}$, the port impedance are renormalized to $Z_0=\SI{15}{\Omega}$. The trend in both IL and BW mirror the measured filter trends in Fig. $\ref{main-Fig_filterResponse}$b and Fig. $\ref{main-Fig_filterResponse}$c, but the IL is under predicted and the BW is still overestimated. These differences are further visualized in Fig. S$\ref{SI_Fig_SimMeas_Filter}$ which compares the measured filter response of device D4 to full finite element simulations of the filter in Ansys HFSS with and without the optimized $N_z$ factors. The higher insertion loss reflects the slightly lower shunt resonator $Q$-factor and higher series resonator resistance as compared to distributed SW resonator models. To improve the IL, the electroplated Au transducers can be made thicker than $\SI{3}{\micro\meter}$ and the Ti wet etch hard mask process in Fig. $\ref{main-Fig_Fab}$i and $\ref{main-Fig_Fab}$p can be optimized for even smoother Au to reduce the filter's resistive loss. Additionally, the shunt resonator transducer width can be decreased sacrificing some spurious mode suppression for high resonance $Q$-factor based on the study in \cite{devitt_distributed_2024}. Alternatively, resonator coupling and filter BW can be scarified by reducing the the length of each resonator to improve IL. The thickness of the GGG membrane beneath each fabricated device is thinner than $\SI{10}{\micro\meter}$ resulting in the higher than designed resonator coupling observed in Fig. S$\ref{SI_measured_resonator_response}$e and smaller than designed filter bandwidth. The influence of the GGG membrane is discussed in Supplementary Information \ref{SI_GGG_Thickness}. As the filter is tuned towards lower center frequencies, eventually the resonator coupling become insufficient to support the filter bandwidth and the passband ripple increases resulting in the sharp drop off in filter bandwidth observed in Fig S$\ref{SI_Fig_Demag}f$. Reducing demagnetization field contrast between the series and shunt resonators (by increasing the shunt YIG width for example) extends the filter's tuning range at the expense of the bandwidth.

\subsection{Influence of GGG Membrane Thickness}
    \label{SI_GGG_Thickness}

    \begin{figure}[p]
        \centering
        \includegraphics[width=\textwidth]{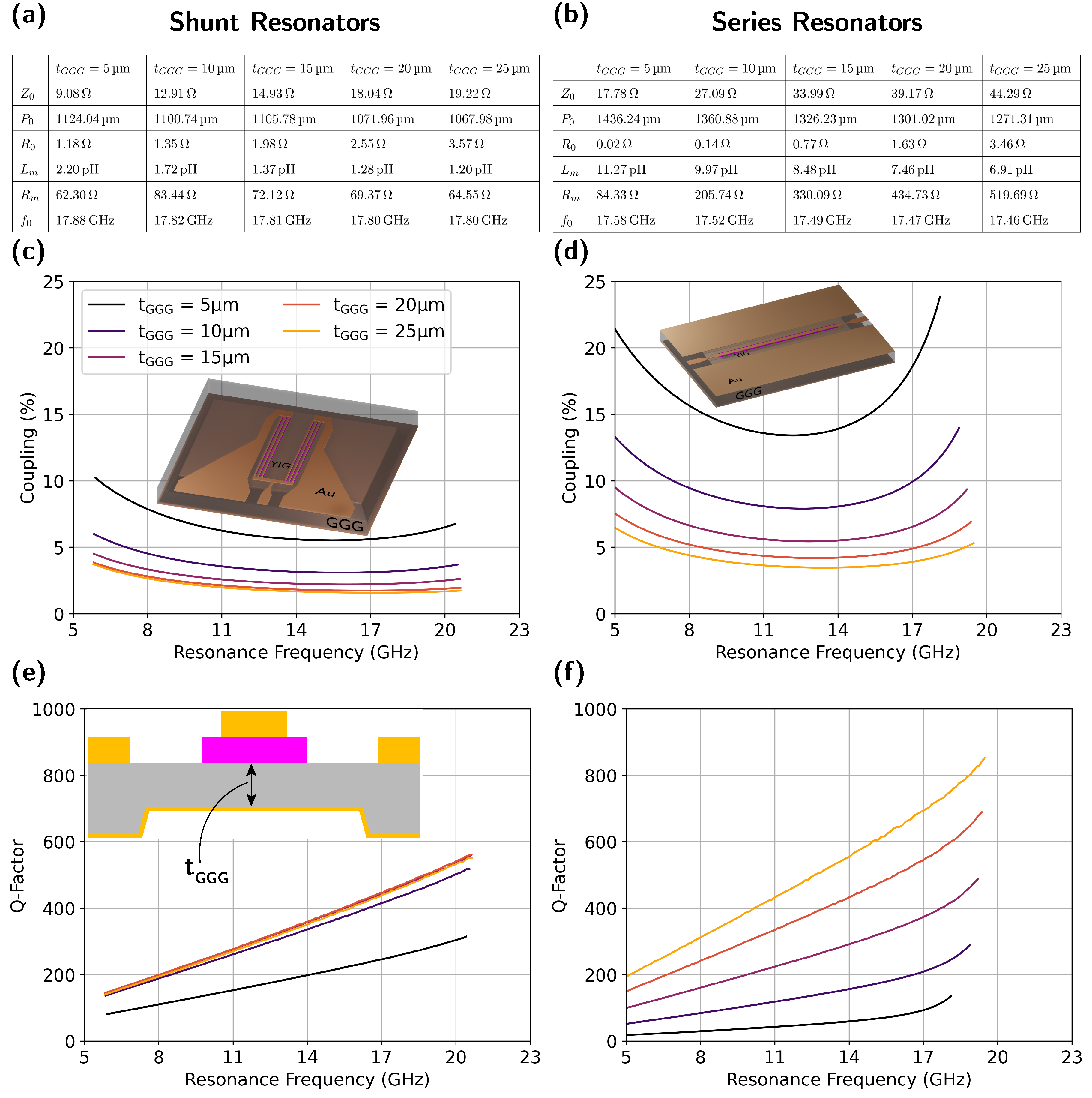}
        \caption{\textbf{Modeled SW hairclip resonator performance over GGG membrane thickness.} The component resonators of device D4 with a $\SI{50}{\micro\meter}$ wide series resonator and $\SI{12}{\micro\meter}$ wide shunt resonators are simulated using a finite element method at $\SI{792.3}{\milli\tesla}$ ignoring demagnetization effects while the thickness of the GGG membrane ($t_{GGG}$) is swept. Each simulated resonator is fit to the distributed SW model from Fig. S$\ref{SI_Fig_Acoustic_SW_compare}$b. \textbf{(a)} and \textbf{(b)} list the fitted model parameters over $t_{GGG}$ for the shunt and series resonators respectively. \textbf{(c)} and \textbf{(d)} show the extracted resonator coupling coefficients while \textbf{(e)} and \textbf{(f)} plot the extracted $\SI{3}{\decibel}$ $Q$-factor over magnetic tuning for the shunt and series resonators respectively.}
        \label{SI_Fig_GGGThickness}
        % \vspace*{-0.1in}
    \end{figure}

    As discussed in \cite{tiwari_high-performance_2025}, the SW resonator coupling and Q-factor strongly depend on the confinement of the transversal RF magnetic field inside the YIG film. For strong coupling, the magnetic field can be tightly confined by placing a ground plane in close proximity to the YIG mesa. In this work and in \cite{tiwari_high-performance_2025}, the GGG substrate is etched from the backside of the chip leaving a thin GGG membrane beneath each YIG mesa and a layer of Au is evaporated over the etched cavity serving as the ground plane. For appreciable improvements in the effective resonator coupling, the thickness of the GGG membrane ($t_{GGG}$) should be below $\SI{20}{\micro\meter}$, however, the resonator $Q$-factor also sharply degrades as $t_{GGG}$ decreases. The SW ladder filter simultaneously requires high $k_{eff}^2$ and high $Q$-factor to realize low-loss and wide-bandwidth performance. As a result, there is an optimal $t_{GGG}$ for a given design specification where $k_{eff}^2$ is large enough to support the desired bandwidth and the $Q$-factor is maximized to give the lowest insertion loss. This optimal membrane thickness is further complicated by the frequency and magnetic bias dependence of the resonator performance. \\
    
\begin{figure}[!t]
        \centering
        \includegraphics[width=\textwidth]{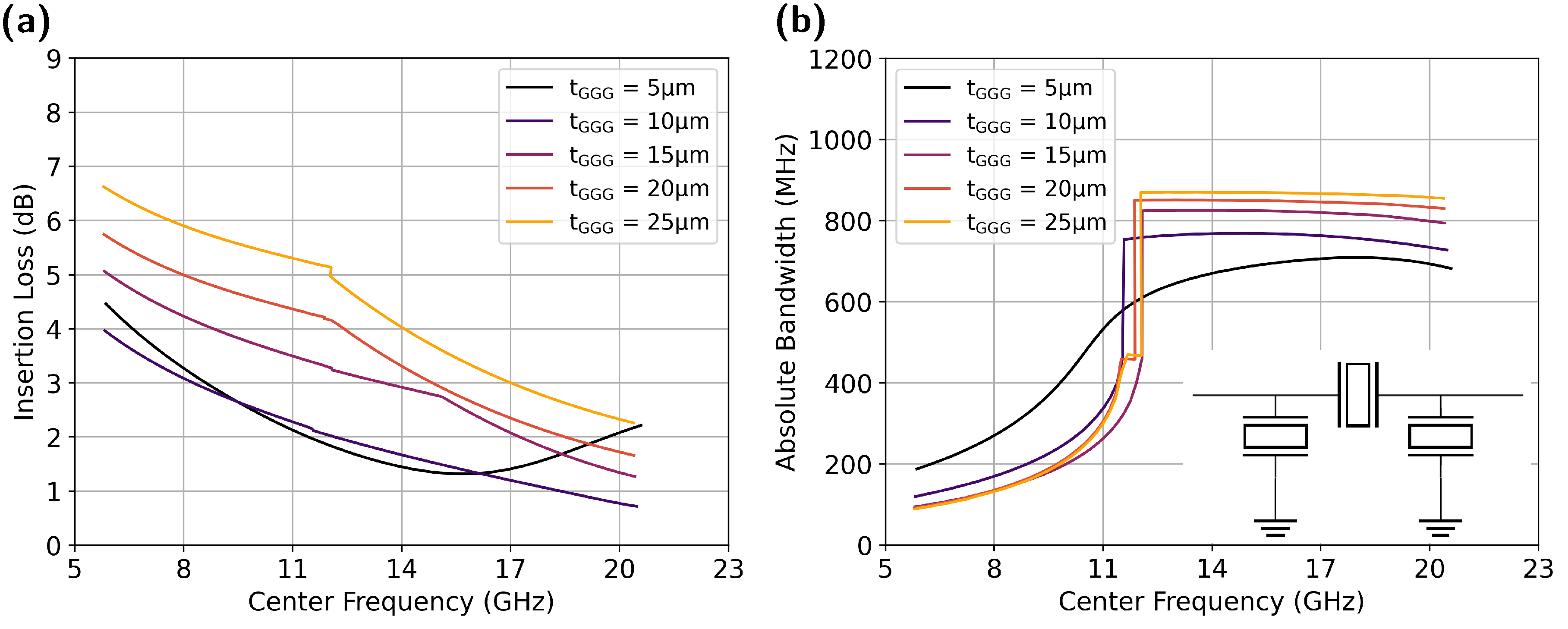}
        \caption{\textbf{Modeled SW ladder filter resonator performance over GGG membrane thickness.} The distributed SW hairclip models from Fig. S$\ref{SI_Fig_GGGThickness}$ are arranged in a $3^{rd}$-order ladder filter with 1 series and 2 shunt resonators using a port impedance of $Z_0=\SI{15}{\ohm}$ and the fitted $N_z$ from Fig. S$\ref{SI_Fig_Demag}$c. \textbf{(a)} and \textbf{(b)} show the resulting ladder filter's simulated insertion loss and bandwidth respectively over magnetic tuning and the GGG membrane thickness.}
        \label{SI_Fig_GGGThickness_Filter}
        % \vspace*{-0.1in}
    \end{figure}

    Fig. S$\ref{SI_Fig_GGGThickness}$ illustrates the frequency dependence of both the series and shunt resonators over the GGG membrane thickness. Both resonator designs are simulated in Ansys HFSS sweeping $t_{GGG}$ from $\SI{5}{\micro\meter}$ to $\SI{25}{\micro\meter}$ at an out-of-plane bias of $H_{DC}=\SI{792.3}{\milli\tesla}$ (ignoring the demagnetizing field) and a distributed SW resonator model is fit to each simulation. The model parameters are listed in Fig. S$\ref{SI_Fig_GGGThickness}$a and S$\ref{SI_Fig_GGGThickness}$b. Using these models, the series and shunt resonator coupling and $Q$-factor are calculated over the applied magnetic bias and over $t_{GGG}$. Consistent with \cite{tiwari_high-performance_2025}, there is a clear trade-off between $Q$-factor and $k_{eff}^2$ and the sensitivity of both parameters increases as the GGG membrane becomes thinner. Fig. S$\ref{SI_Fig_GGGThickness_Filter}$ extends the resonator performance trade-off to filter performance. The resonator models are configured in a 3\textsuperscript{rd}-order 1-series, 2-shunt ladder filter configuration with a port impedance of $Z_0=\SI{15}{\ohm}$. A correction factor is applied to each model's resonance frequency for the demagnetization field using the fitted $N_z$ from Fig. S$\ref{SI_Fig_Demag}$c. At large $t_{GGG}$, the filters show high insertion loss due to the insufficient resonator $k_{eff}^2$. Conversely, the filter bandwidth decreases as $t_{GGG}$ decreases primarily due to the increased series $k_{eff}^2$. Therefore the discrepancy between the measured and simulated filter bandwidth after correcting for $N_z$ in Fig. S$\ref{SI_Fig_SimMeas_Filter}$b can be attributed to a slight GGG over-etch in the timed hot phosphoric acid from Fig. $\ref{main-Fig_Fab}$o. The wax bonding and thinning process in Fig. $\ref{main-Fig_Fab}$l introduces a $\pm \SI{1}{\micro\meter}$ $t_{GGG}$ nonuniformity across the $\SI{15}{\milli\meter}\times\SI{15}{\milli\meter}$ chip which also influences the resonator performance.
    
    % \clearpage

\section{Supplementary Information: 5\textsuperscript{th}-Order Filters}
	\label{5th_order_filter}

\begin{figure}[!t]
        \centering
        \includegraphics[width=\textwidth]{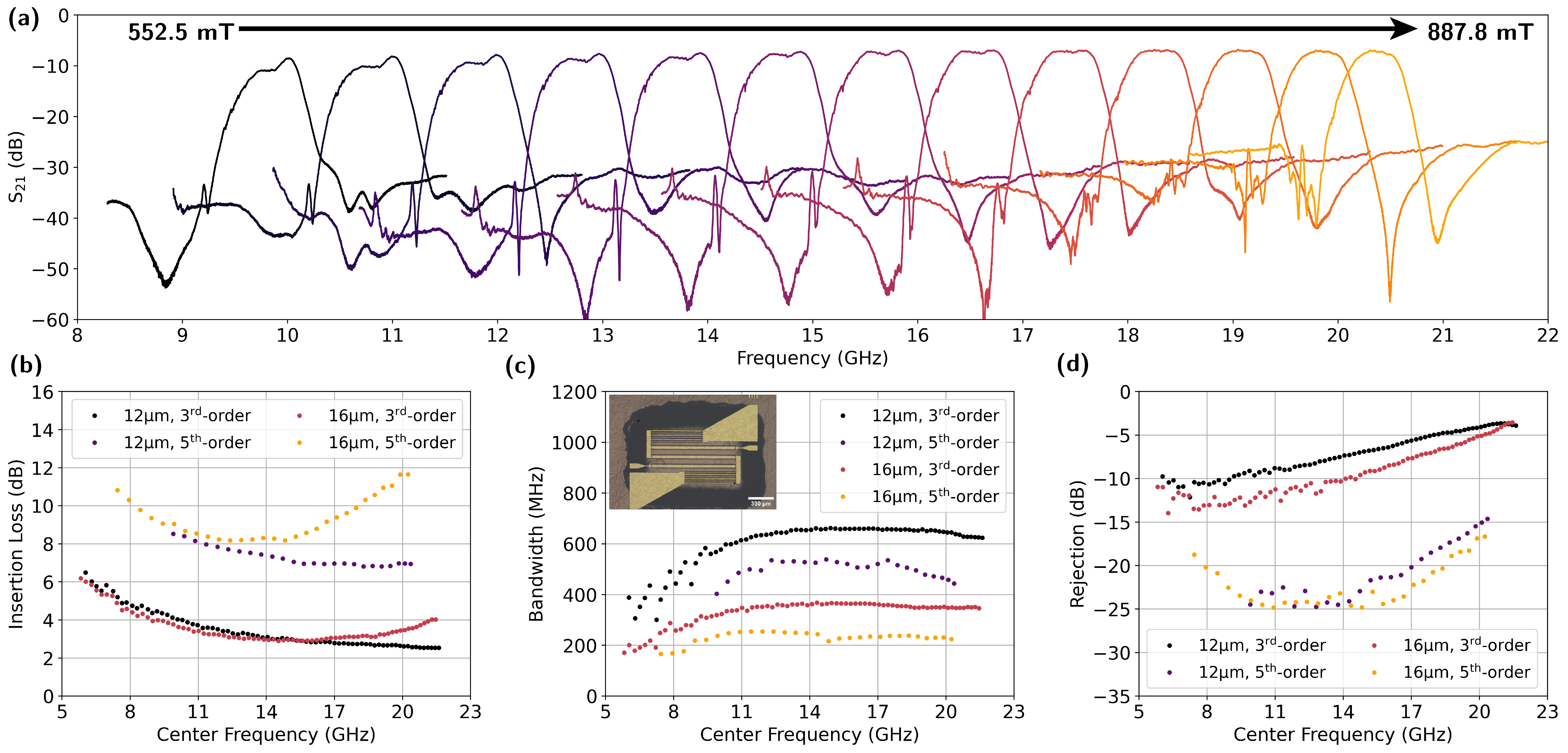}
        \caption{\textbf{Measured Frequency Response of 5\textsuperscript{th}-Order SW Ladder Filters.} The higher order filters shown in Fig. \ref{main-Fig_Design}f are comprised of 3 series and 2 shunt SW hairclip resonators. \textbf{(a)} Tuning response of filter E4 from $\SI{552.5}{\milli\tesla}$ to $\SI{887.8}{\milli\tesla}$. The response is cropped around the passband for clarity. \textbf{(b)} Insertion loss, \textbf{(c)} bandwidth, and \textbf{(d)} rejection comparison between 3\textsuperscript{rd}-order and 5\textsuperscript{th}-order SW Ladder Filters.}
        \label{SI_Fig_5poleFilter}
        % \vspace*{-0.1in}
    \end{figure}

    In addition to 3\textsuperscript{rd}-order filters as discussed in the matin text, the single bias SW ladder filter concept can be extended to higher order filters to provide significantly enhanced rejection at the expense of insertion loss and bandwidth by including additional resonator elements as is often done in with acoustic filters in the RF MEMS community \cite{anusorn_frequency_2025}. To demonstrate this for the SW ladder filters, a 5\textsuperscript{th}-order filter (shown in Fig. \ref{main-Fig_Design}f) is designed and fabricated consisting of 3 series resonators and 2 shunt resonators. The component resonators are identical to those used in the 3\textsuperscript{rd}-order filters, except the YIG fins in the shunt resonator are combined into 6 element arrays. Relative to the 3\textsuperscript{rd}-order filter, the inclusion of additional shunt resonators would reduce the filter's impedance so two series resonators are added instead to maintain the filter impedance match closer to $Z_0=\SI{50}{\ohm}$. However, the data reported in Fig. S$\ref{SI_Fig_5poleFilter}$ is renormalized to $\SI{15}{\ohm}$ for direct comparison with the 3\textsuperscript{rd}-order filters. The geometry of the 5\textsuperscript{th}-order filter is optimized through finite element simulation for the highest tuning range and the most compact YIG resonator layout. A smaller filter area helps mitigate detrimental effects from a non-uniform out-of-plane magnetic field and reduces the size of the thin GGG membrane, improving fabrication yield. The three, closely-spaced parallel series resonators act as coupled transmission lines reducing the electrical length of the filter which extends the maximum tuning range. Fig. S$\ref{SI_Fig_5poleFilter}$a shows the frequency response of device E4 over magnetic field.  Fig. S$\ref{SI_Fig_5poleFilter}$b-d show the measured IL, BW, and rejection for two 5\textsuperscript{th}-order filters with shunt YIG widths of $\SI{12}{\micro\meter}$ and $\SI{16}{\micro\meter}$ compared 3\textsuperscript{rd}-order filters with the same YIG dimensions. The insertion loss of the filter at higher frequencies is primarily limited by the resistive loss through the series resonators so the 5\textsuperscript{th}-order filter has approximately triple the insertion loss. At low frequencies, the YIG exhibit low $Q$-factors and the resonators have insufficient coupling to support the high filter bandwidth leading to degraded insertion loss mirroring the trend in the 3\textsuperscript{rd}-order filter performance. The rejection shows over a $\SI{10}{\decibel}$ improvement compared to the 3\textsuperscript{rd}-order filters.

    \vfill
    \newpage
    
\section{Supplementary Information: Filter Nonlinearity and Power Handling}
    \label{Si_sec_nonlinearity}
    \begin{figure}[!t]
        \centering
        \includegraphics[width=\textwidth]{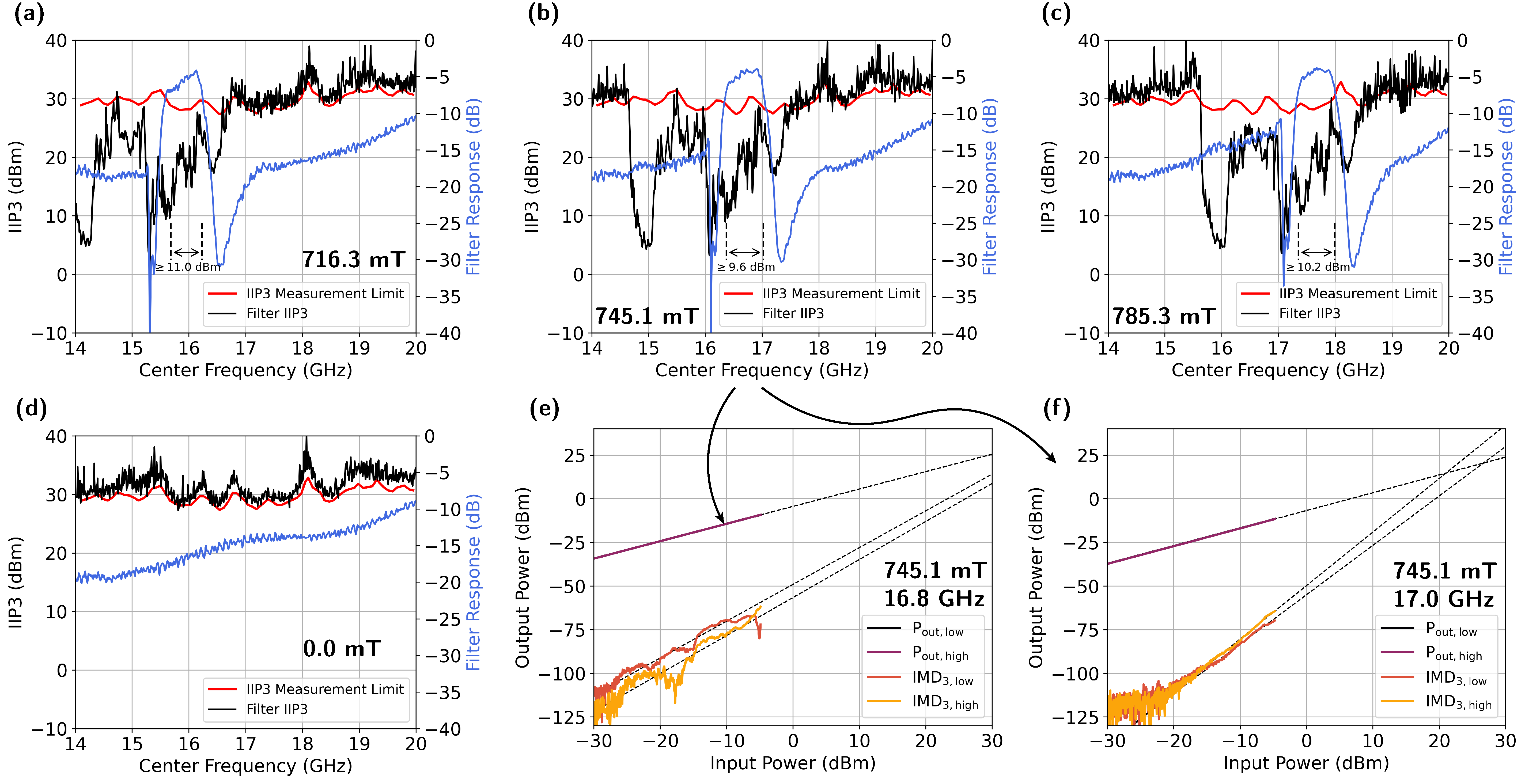}
        \caption{\textbf{Measured SW ladder filter IIP3.} The filter IIP3 of device D4 is evaluated over frequency, power, and magnetic field. $\textbf{(a)-(d)}$ show the measured IIP3 at $\SI{716.3}{\milli\tesla}$, $\SI{745.1}{\milli\tesla}$, $\SI{785.3}{\milli\tesla}$, and $\SI{0}{\milli\tesla}$ as the center frequency of the two input tones are swept from $\SI{14}{\giga\hertz}$ to $\SI{20}{\giga\hertz}$ with a constant input power of $\SI{-10}{dBm}$ and tone spacing of $\SI{1}{\mega\hertz}$. The measured IIP3 for a calibration thru shown in red determines the measurement sensitivity limit. The filter response in blue is the average output power of the two fundamental tones normalized by the average input tone power. \textbf{(e)} and \textbf{(f)} show the fundamental and third-order intermodulation tone powers as the input power is swept from $\SI{-30}{dBM}$ to $\SI{-5}{dBM}$ with a $\SI{1}{\mega\hertz}$ tone spacing at a fixed magnetic bias of $\SI{745.1}{\milli\tesla}$. The center frequency of the fundamental tone is \textbf{(e)} $\SI{16.8}{\giga\hertz}$ and \textbf{(f)} $\SI{17.0}{\giga\hertz}$.}
        \label{SI_Fig_Nonlinearity}
        % \vspace*{-0.1in}
    \end{figure}

    \begin{figure}[!b]
        \centering
        \includegraphics[width=\textwidth]{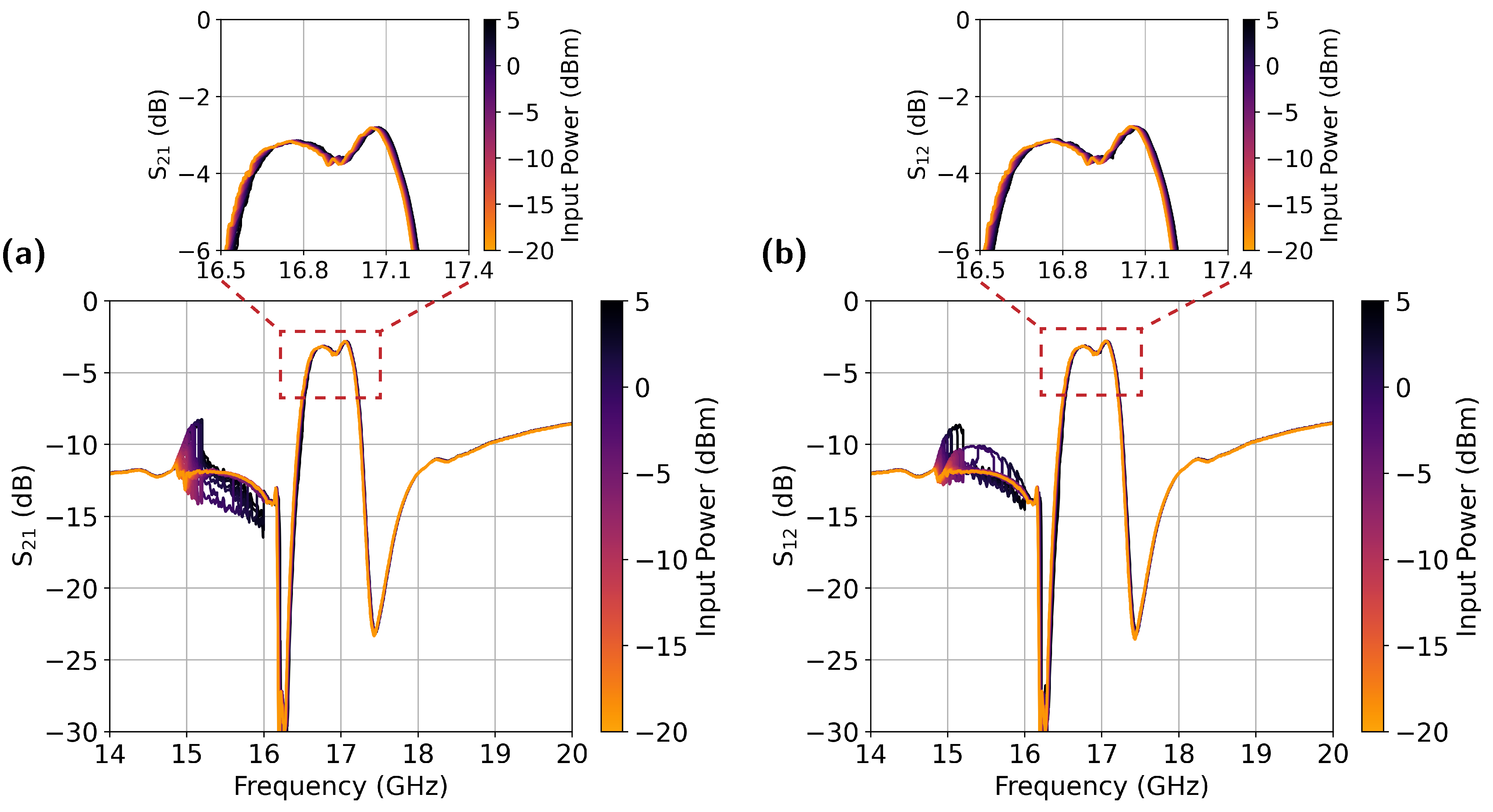}
        \caption{\textbf{SW ladder filter performance over input power.} \textbf{(a)-(b)} Measured SW ladder filter response of device D4 as the input power is sweep from $\SI{-20}{dBm}$ to $\SI{5}{dBm}$. Within the $\SI{3}{\decibel}$ passband, no significant power dependent response is observed. The filter center frequency drifts higher during the measurement as the electromagnet providing the uniform out-of-plane bias heats the filter chip.}
        \label{SI_Fig_PowerHandling}
        % \vspace*{-0.1in}
    \end{figure}
    
    The filter linearity and power handling are evaluated using device D4 from Fig. $\ref{main-Fig_Design}e$. The physical experimental setup is unchanged from Supplementary Information \ref{Measurement_Setup}. For the intermodulation distortion measurements in Fig. S$\ref{SI_Fig_Nonlinearity}$, a PNA-X (N5244B) with two internal sources and an internal combining network is used. In Fig. S$\ref{SI_Fig_Nonlinearity}$a-d, the PNA-X generates two tones ($f_1,f_2$) at a fixed $-10\text{ dBm}$ output power at port 1 with a spacing of $f_2-f_1=\SI{1}{\mega\hertz}$ while sweeping the center frequency of the two tones from $\SI{14}{\giga\hertz}$ to $\SI{20}{\giga\hertz}$ with $501$ points. At port 2, the PNA-X measures the power at the fundamental tone frequencies with an IFBW of $\SI{30}{\hertz}$ and at the 3\textsuperscript{rd}-order intermodulation product (IMD3) frequencies ($2f_1 - f_2$ and $2f_2-f_1$) with an IFBW of $\SI{5}{\hertz}$. A power calibration is performed over frequency at the end of the \SI{2.92}{\milli\meter} cable connected to the non-magnetic GSG probes. Additionally, a SOLT calibration is performed over frequency at the end of the GSG probes. At each center frequency, the input referred 3\textsuperscript{rd}-order intercept point  (IIP3) is calculated assuming the power in each IMD3 tone is proportional the cube of the input power ($P_{IMD3}\propto P_{in}^3$) using
    \begin{equation}
        IIP3 = P_{in} + \frac{1}{2}\left(P_{out}-P_{IMD3}\right).
    \end{equation}
    The IIP3 reported in S$\ref{SI_Fig_Nonlinearity}$a-d is the average of the high tone and low tone IIP3 calculation. In Fig. S$\ref{SI_Fig_Nonlinearity}$e-f, the input power is swept from $-30 \text{ dBm}$ to $-5 \text{ dBm}$ using $601$ points with a bias of $\SI{745.1}{\milli\tesla}$ and fixed center frequencies of $\SI{16.8}{\giga\hertz}$ and $\SI{17.0}{\giga\hertz}$. All other measurement settings are the unchanged.\\

    Prior to measuring the SW ladder filter, the linearity of a calibration thru standard (CSR-8 calibration substrate) is measured and the IIP3 is plotted in red in Fig. S\ref{SI_Fig_Nonlinearity}a-d. The thru is linear and the power of the intermodulation products is below the noise floor so this line represents the IIP3 measurement limit in this experimental setup. Fig. S$\ref{SI_Fig_Nonlinearity}$a-d shows the measured IIP3 with a magnetic bias of $\SI{716.3}{\milli\tesla}$, $\SI{745.1}{\milli\tesla}$, $\SI{785.3}{\milli\tesla}$, and $\SI{0}{\milli\tesla}$ alongside the normalized filter response ($P_{out, avrg}/P_{in, avrg}$). At $\SI{0}{\milli\tesla}$, there is no spinwave resonance within the measurement frequency, and the filter linearity increases beyond the measurement limit. The filter also shows high linearity while biased with a minimum in-band IIP3 of $11.0 \text{ dBm}$, $9.6 \text{ dBm}$, and $10.2 \text{ dBm}$ at $\SI{716.3}{\milli\tesla}$, $\SI{745.1}{\milli\tesla}$, and $\SI{785.3}{\milli\tesla}$ respectively. In the lower stopband, a strong nonlinear response is observed below the resonance frequency of the series resonator. By overlaying the measured resonator responses in Fig. S$\ref{SI_measured_resonator_response}$b-c with the IIP3 measurements, this nonlinearity is associated with low-$Q$ spurious spinwave modes in the shunt resonators.\\
   
    \begin{figure}[!t]
        \centering
        \includegraphics[width=0.5\textwidth]{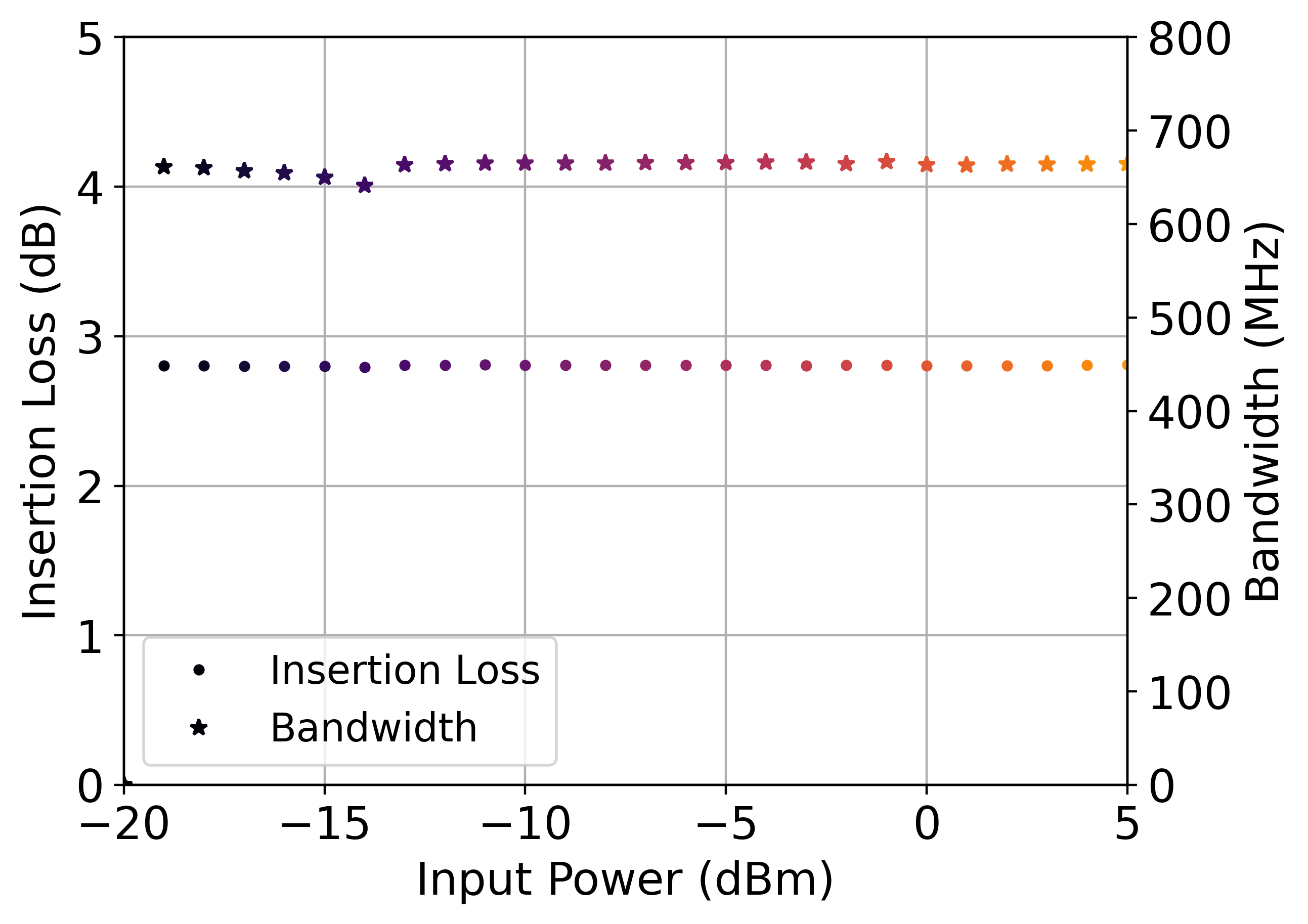}
        \caption{\textbf{SW ladder filter insertion loss and bandwidth over input power.} The measured SW ladder filter insertion loss and bandwidth of device D4 as the input power is sweep from $\SI{-20}{dBm}$ to $\SI{5}{dBm}$. The filter's probe pads are de-embedded from the measurement and the port impedance is renormalized to $Z_0=\SI{15}{\ohm}$.}
        \label{S12_PowerSweep_IL_BW}
        % \vspace*{-0.1in}
    \end{figure}
    
    Fig. S$\ref{SI_Fig_PowerHandling}$ shows the filter s-parameters of device D4 as the input power is swept from $-20 \text{ dBm}$ to $5 \text{ dBm}$. The probe pads are de-embedded from each measurement and port impedance is renormalized to $Z_0=\SI{15}{\ohm}$. Within the filter's $\SI{3}{\decibel}$ bandwidth, no power dependent performance is observed as indicated in Fig. S$\ref{S12_PowerSweep_IL_BW}$. Due to the YIG resonator's uncompensated temperature coefficient of frequency (TCF), the filter's center frequency drifts towards higher frequencies during the experiment as the electromagnet heats the filter chip. Below the passband near the shunt resonator's low-$Q$ spurious SW modes, a strong power dependence is observed in Fig. S$\ref{SI_Fig_Nonlinearity}$. With increasing power, the filter rejection degrades. 
    With further study, this nonlinearity could be used to provide an SNR-enhancing effect.

    \clearpage
    \newpage
    
    \section{Supplementary Information: Frequency Agile Radio}
    \label{Si_sec_radio}

    \begin{figure}[!b]
        \centering
        \includegraphics[width=\textwidth]{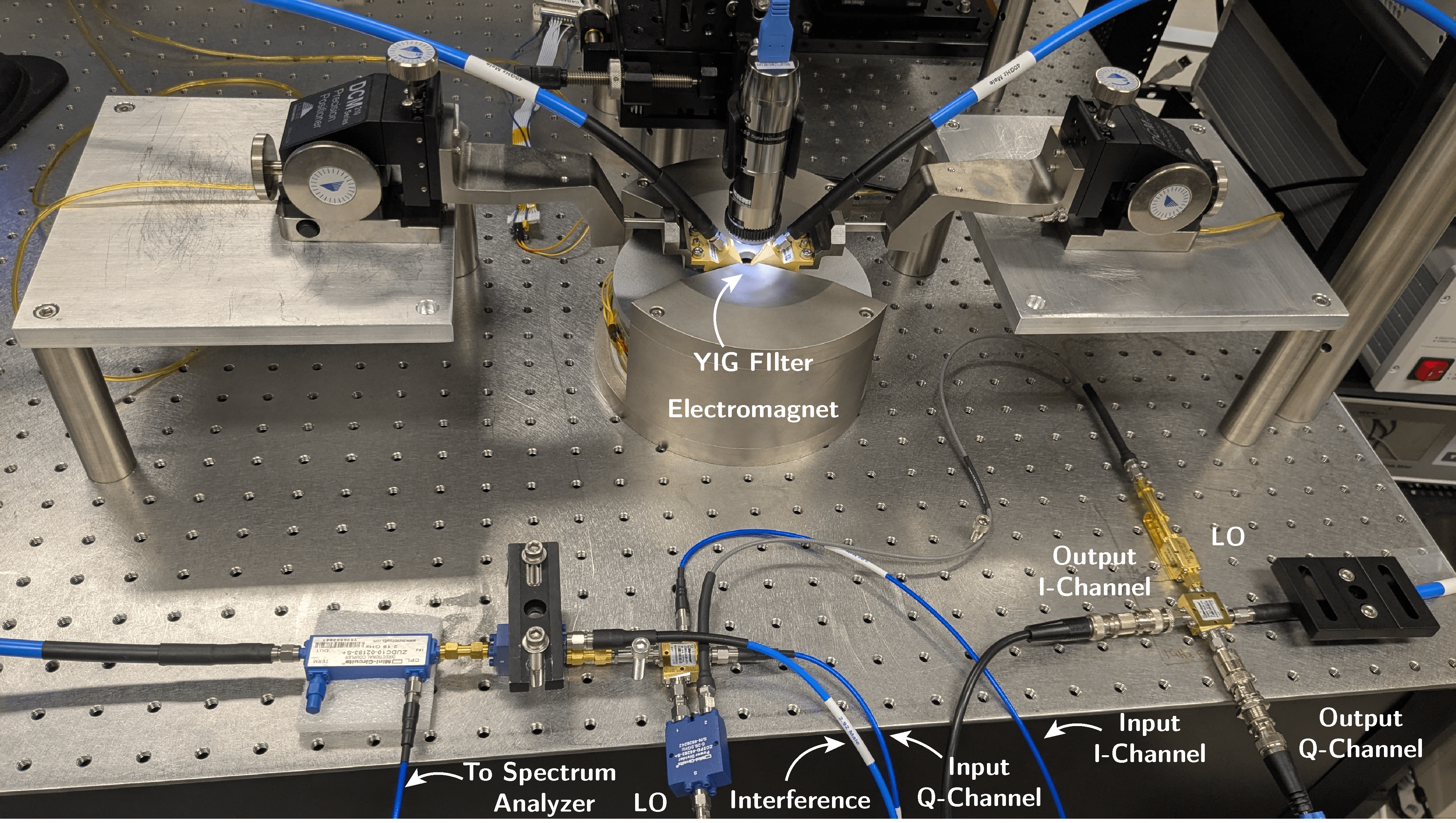}
        \caption{\textbf{Photograph of the SW ladder filter frequency agile radio experimental setup.} }
        \label{SI_Fig_radio_setup_annotated}
        % \vspace*{-0.1in}
    \end{figure}
    
\begin{figure}[!t]
        \centering
        \includegraphics[width=\textwidth]{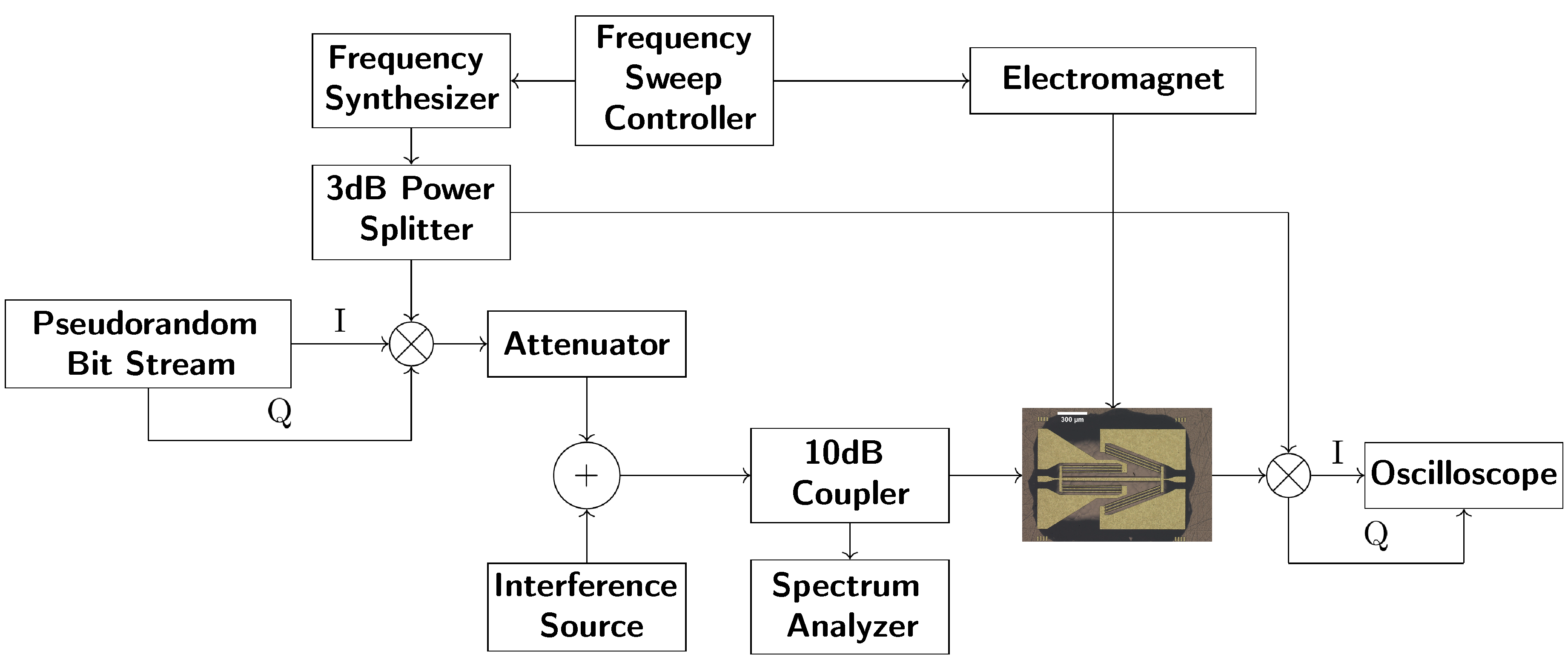}
        \caption{\textbf{Functional block diagram of the frequency tuneable radio with spectrum monitoring.}}
        \label{SI_Fig_Radio_Jamming}
        % \vspace*{-0.1in}
    \end{figure}

    Fig. S\ref{SI_Fig_radio_setup_annotated} shows a photograph of the experimental setup for the frequency agile radio. A Keysight 33500B waveform generator is used to generate the pseudorandom bit streams at $10 \text{ Mbps}$ for the in-phase (I) and quadrature (Q) channels with a $\SI{400}{\milli\volt}$ amplitude. An Agilent E8257D provides the local oscillator (LO) signal for the ZMIQ-243H-K+ IQ mixers and a ZC2PD-06263-S+ $\SI{3}{\decibel}$ splitter divides the LO between the the transmitter (Tx) and receiver (Rx). A wideband $\SI{20}{\decibel}$ attenuator is placed at the output of the Tx IQ mixer to simulate an additive white Gaussian noise (AWGN) channel. The probe, filter, and electromagnet setup is identical to Fig. S\ref{SI_Fig_Measurement_setup} and device B4 from Fig. S\ref{SI_Fig_device_labels} is used. A Tektronix MS046 oscilloscope records the demodulated IQ data stream. A second E8257D provides the Gaussian noise amplitude modulated interference which is combined with the transmitted signal using a second  ZC2PD-06263-S+ splitter. A \SI{10}{\decibel} coupler (ZUDC10-02183-S+) and an Agilent PXA signal analyzer inserted between the splitter and filter input (Fig. S\ref{SI_Fig_Radio_Jamming}) are used to monitor the received spectrum and generate the plot in Fig. \ref{main-Fig_radio}e.\\

    The radio demonstration in Fig. S$\ref{SI_Fig_radio_setup_annotated}$ uses quadrature amplitude modulation (QAM) where information is encoded in both the phase and amplitude of the carrier \cite{proakis_fundamentals_2005}. QAM is a coherent modulation technique where the carrier's phase must be accurately tracked at the receiver. If the receiver's LO is not frequency and phase locked to the transmitted carrier, then the information in the phase of the QAM constellation is lost. From Fig. $\ref{main-Fig_radio}a$, the transmitter and receiver share the same LO, so they are frequency locked. However, the group delay from the Tx output to the receiver's IQ mixer differs from the group delay in the LO's path from the splitter output to receiver's IQ mixer. As a result, the carrier and receiver LO have a frequency-dependent phase difference. This problem can be observed in the supplementary videos where the demodulated constellations rotate as the carrier frequency is swept. By including a phase locked locked loop (PLL) at the receiver tracking $4f_0$, the receiver's LO can be phase locked to the carrier as described in \cite{proakis_fundamentals_2005}. Generally, frequency hopped spread spectrum (FHSS) radios do not use coherent modulation techniques, however, because maintaining phase coherence during the the rapid carrier frequency shifts and over a channel with frequency selective fading and Doppler shifts is challenging \cite{proakis_fundamentals_2005, li_analysis_2021, torrieri_principles_2018}. Instead, they employ noncoherent demodulation techniques where the knowledge of the carrier's phase is not required. FHSS systems still require frequency locking to the carrier and use multiple synchronization circuits to track the carrier's frequency over time.\\

    % The performance of the YIG filter in the radio system is also evaluated when an out-of-band interference signal is present and Fig. S\ref{SI_Fig_Radio_Jamming}a illustrates the experimental setup. A strong amplitude modulated gaussian noise tone is introduced outside the YIG filter passband and the resulting power spectrum at the output of a $\SI{10}{\decibel}$ coupler is shown in Fig. S\ref{SI_Fig_Radio_Jamming}b illustrating the relative signal and interference levels at the input of the bandpass filter. Fig. S\ref{SI_Fig_Radio_Jamming}c--d show the measured demodulated IQ constellations at a data rate of $20 \text{ Mbps}$ with and without interference. Despite the strong out-of-band interference, the YIG filter provides sufficient attenuation to preserve low bit error rates for the signal of interest. A video is also provided in the supplementary material where the interference signal is swept across frequency while $f_0$ and the YIG bias are fixed. 
    
    % \clearpage

	\bibliographystyle{IEEEtran}
	\bibliography{ARROW_bib}